\documentclass[
reprint,
prd,
twocolumn,
nofootinbib,
tightenlines %
floatfix,
 amsmath,
showpacs ,amssymb, aps,% linenumber
superscriptaddress
]{revtex4-2}
\usepackage{amsmath,slashed}
\usepackage[left]{lineno}
\usepackage{units}
\usepackage{comment}
\usepackage{graphicx}
\usepackage{amsmath,amssymb}
\usepackage{support-caption}
\usepackage{subcaption}
\usepackage{bm}
\usepackage[shortlabels]{enumitem}
\usepackage{float}
\usepackage{soul}
\usepackage{natbib}
\usepackage{wasysym}
\usepackage{aas_macros}
\graphicspath{{images/}}
\RequirePackage{lineno}
\usepackage[usenames,dvipsnames,table]{xcolor}
\usepackage{appendix}
\usepackage{extdash}
\usepackage{colortbl}

\definecolor{lightblue}{HTML}{1F77B4}

\newcommand{\SPA}{School of Physics and Astronomy, Monash University, Vic 3800, Australia}
\newcommand{\OzGravMonash}{OzGrav: The ARC Centre of Excellence for Gravitational Wave Discovery, Clayton VIC 3800, Australia}
\newcommand{\AstroThreeD}{ARC Centre of Excellence for All Sky Astrophysics in 3 Dimensions (ASTRO 3D), Canberra, ACT 2611, Australia}
\newcommand{\GeorgiaTech}{Center for Relativistic Astrophysics and School of Physics, Georgia Institute of Technology, Atlanta, Georgia 30332, USA}
\newcommand{\Flatiron}{Center for Computational Astrophysics, Flatiron Institute, 162 5th Ave, New York, NY 10010}
\definecolor{Waveform1}{HTML}{1f77b4}
\definecolor{Waveform2}{HTML}{ff7f0e}
\definecolor{Waveform3}{HTML}{2ca02c}
\definecolor{Waveform4}{HTML}{d62728}
\definecolor{Waveform5}{HTML}{9467bd}
\definecolor{Waveform6}{HTML}{8c564b}
\definecolor{Waveform7}{HTML}{e377c2}
\definecolor{Waveform8}{HTML}{7f7f7f}
\definecolor{Waveform9}{HTML}{bcbd22}

\begin{document}

\title{Detection and parameter estimation of binary neutron star merger remnants}

\author{Paul J. Easter}
\email{paul.easter@monash.edu}
\affiliation{\SPA}
\affiliation{\OzGravMonash}
\author{Sudarshan Ghonge}
\affiliation{\GeorgiaTech}
\author{Paul D. Lasky}
\affiliation{\SPA}
\affiliation{\OzGravMonash}
\author{Andrew R. Casey}
\affiliation{\SPA}
\affiliation{\AstroThreeD}
\author{James A. Clark}
\affiliation{\GeorgiaTech}
\author{Francisco Hernandez Vivanco}
\affiliation{\SPA}
\affiliation{\OzGravMonash}
\author{Katerina Chatziioannou}
\affiliation{\Flatiron}
\pacs{
}

\begin{abstract}
    Detection and parameter estimation of binary neutron star merger remnants can shed light on the physics of hot matter at supranuclear densities.
    Here we develop a fast, simple model that can generate gravitational waveforms, and show it can be used for both detection and parameter estimation of post-merger remnants.
    The model consists of three exponentially-damped sinusoids with a linear frequency-drift term.
    The median fitting factors between the model waveforms and numerical-relativity simulations exceed 0.90.
    We detect remnants at a post-merger signal-to-noise ratio of $\ge 7$ using a Bayes-factor detection statistic with a threshold of 3000.
    We can constrain the primary post-merger frequency to $\pm_{1.2}^{1.4}\%$ at post-merger signal-to-noise ratios of 15 with an increase in precision to $\pm_{0.2}^{0.3}\%$ for post-merger signal-to-noise ratios of 50.
    The tidal coupling constant can be constrained to $\pm^{9}_{12}\%$ at post-merger signal-to-noise ratios of 15, and $\pm 5\%$ at post-merger signal-to-noise ratios of 50 using a hierarchical inference model.
\end{abstract}
\maketitle
    \section{Introduction}
    \label{sec:introduction}
    Gravitational waves have been directly detected from the inspiral of binary neutron star mergers~\cite{GW170817Detection,GW190425Detection}. 
    The  post-merger remnant may promptly collapse into a black hole, or form a hot, differentially-rotating neutron star~\cite{Baumgarte2000,shibata2000bar}, which emits gravitational waves~\cite[e.g.][]{Nakamura1994,New1997,Rasio1999,Shibata2005}.
    Numerical-relativity simulations of post-merger remnants show relationships between the gravitational-wave spectra and a number of progenitor properties through quasi-universal relationships~\cite[e.g.][]{Bauswein2012,Bauswein2012a,Hotokezaka2013,Bernuzzi2014,Takami2014,Bernuzzi2015,Bauswein2015,Takami2015,Rezzolla2016,Bauswein2019}.
    Of particular interest is the relationship between the progenitor tidal coupling constant and the primary post-merger oscillation frequency for baryonic equations of state~\cite{Takami2015,Bernuzzi2015,Rezzolla2016}, which can be used to place constraints on the tidal coupling constant.\par
    
    Gravitational-wave spectra generated from numerical-relativity simulations show consistent  features related to the dynamics of the surviving remnant.
    A dominant peak, designated as $f_{\mathrm{peak}}$~\cite[][]{Oechslin2007a}, is produced by the fundamental oscillations of the bar-mode deformed post-merger remnant~\cite[e.g.][]{Shibata2003a,Rezzolla2010,Giacomazzo2011,Hotokezaka2011,Bauswein2012}.
    The frequencies of four possible peaks can be labelled as $(f_{2-0}, f_{\mathrm{spiral}},f_{\mathrm{peak}},f_{2+0})$ in ascending order~\cite{Bauswein2015}. 
    The peaks at frequencies $f_{2-0},f_{2+0}$ may result from coupling between a spherically-symmetrical quasi-radial oscillation mode and $f_{\mathrm{peak}}$~\cite{Stergioulas2011}.
    The peak at frequency $f_{\mathrm{spiral}}$ may result from the slower rotation-rate of tidally-deformed matter at the outer edges of the post-merger remnant~\cite{Bauswein2015}.
    See~\cite{Takami2014,Takami2015} for an alternative proposed explanation of the frequency peaks.
\par
\begin{figure*}
        \centering
        \includegraphics[scale=0.5]{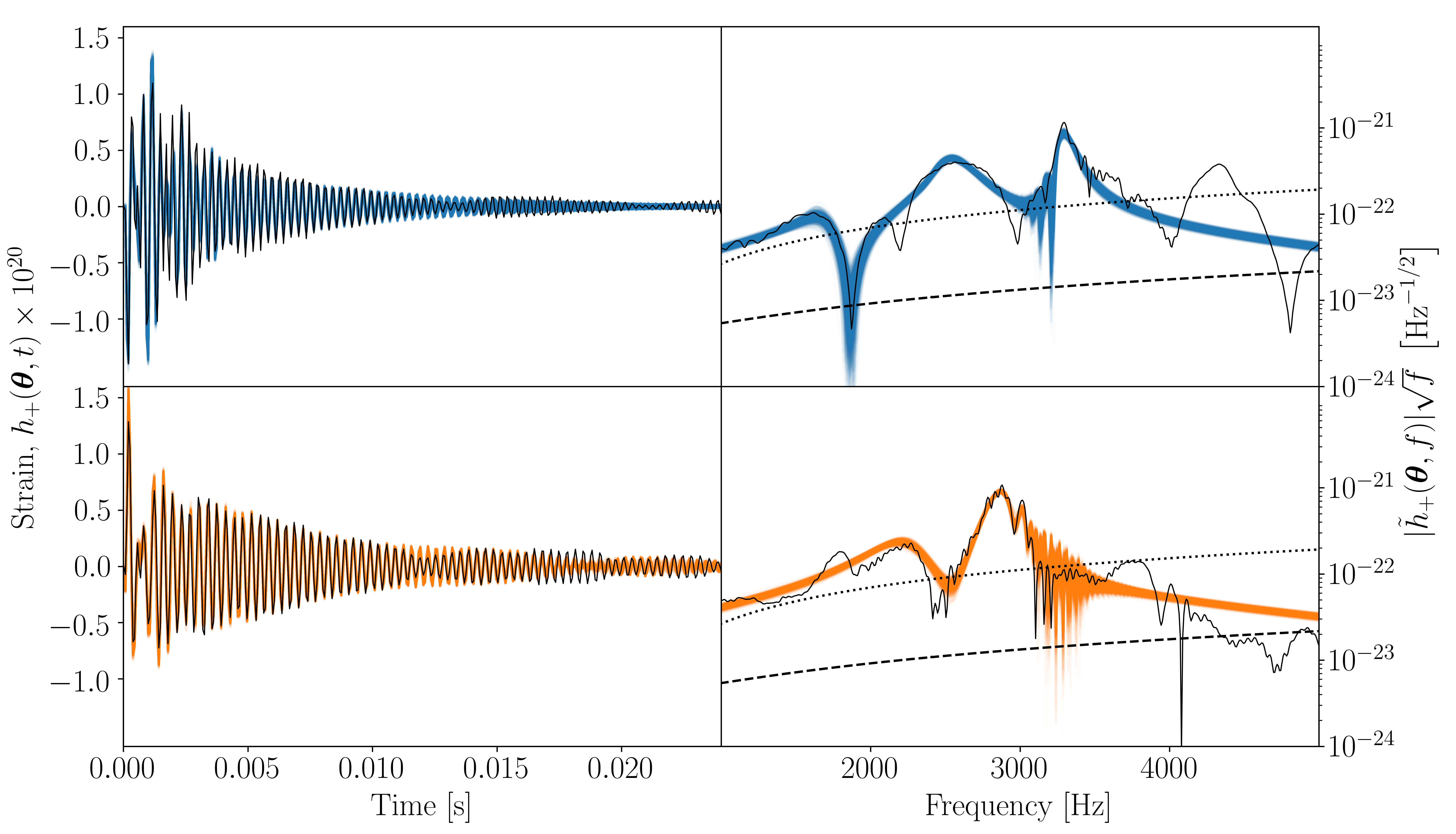}
        \caption
        {Waveform reconstruction of numerical-relativity post-merger signal injections. Top panels: time (left) and frequency (right) domain reconstructions of a numerical-relativity simulation using the SLy equation of state with equal-mass, $1.35\,\mathrm{M}_\odot$, neutron stars (waveform SLy-M1.350-$\Lambda$390). The post-merger waveform (black curve) is injected at a post-merger signal-to-noise ratio of 50. The reconstructed waveforms are shown in blue. Bottom panels: same as the top panels except the injected waveform is using the LS220 equation of state with equal mass, $1.35\,\mathrm{M}_\odot$, neutron stars (waveform LS220-M1.350-$\Lambda$684). The reconstructed waveforms are shown in orange. Noise sensitivity curves are shown for Advanced LIGO (dashed black) and Advanced Virgo (dotted black) for plots on the right.} 
        \label{fig:TimeFrequencyResponse}
\end{figure*}
    In this paper, we develop a Bayesian detection and parameter-inference pipeline.
    Normally these pipelines require a large bank of waveforms.
    Numerical-relativity simulations cannot be used to generate these waveforms as each simulation requires~${\sim\,\mathcal{O}(10^5)}$ CPU hours to complete~\cite{Takami2015}.
    We develop a fast, simple model of gravitational waves for post-merger remnants that phenomenologically incorporates the main frequencies previously mentioned.
    Our model produces waveforms in a time-frame that is suitable for use in detection and parameter estimation of binary neutron star post-merger remnants.\par
    
    We match numerical-relativity waveforms with fitting factors of 0.92-0.97.
    This model addresses the two restrictions that prevent matched filtering of post-merger gravitational-wave strain: computational time and poor fitting-factors.
    The waveforms are derived from a hybrid of the two models outlined in~\cite{Bauswein2016} and~\cite{Bose2018}. 
    Our model is agnostic to the locations of the frequency peaks and uses Bayesian statistics to determine the actual peak frequencies.
    Furthermore, the addition of a frequency drift term allows for secular changes in the frequency peak locations.
    With post-merger signal-to-noise ratios of $\ge 15$, the model can localise the primary post-merger frequency to $\pm_{1.2}^{1.4}\%$ at 95\% confidence, reducing to $\pm_{0.2}^{0.3}\%$ at post-merger signal-to-noise ratio of 50.
    Using the hierarchical model developed in~\cite{Easter2019} we can then constrain the tidal parameters and compactness of the progenitor neutron stars.
    The tidal coupling constant is constrained to $\pm_{12}^{9}\%$ at post-merger signal-to-noise ratios of 15 for a 95\% confidence interval. 
    At post-merger signal-to-noise ratios of 50 this tightens to $\pm 5 \%$. \par
    
    In Section~\ref{sec:methodology} we outline the model and associated methods used in this paper. 
    In Section~\ref{sec:modelfit} we validate the model fits in the time and frequency domains and quantify the goodness of the fits.
    In Section~\ref{sec:sensitivity} we use a Bayes factor detection statistic to determine at what post-merger signal-to-noise ratios a detection occurs and test how the model performs due to uncertainty in the inspiral coalescence time.
    In Section~\ref{sec:ParameterEstimation} we calculate posteriors of the dominant post-merger frequency and introduce the hierarchical model from~\cite{Easter2019} to find the equation of state parameters for the progenitors.
    We find constraints on both the tidal coupling constant and the compactness of the progenitors.

    \section{Methodology}\label{sec:methodology}
    We adopt a model for the post-merger gravitational-wave signal consisting of three exponentially damped sinusoids~\cite{Bauswein2016} with additional linear frequency drift terms~\cite{Bose2018}. 
    The plus, $h_{+}(\boldsymbol{\theta},t)$, polarisation of the gravitational-wave strain is extracted from the right circular polarisation, $h(\boldsymbol{\theta},t)$,  as follows:
\begin{align}
        h(\boldsymbol{\theta},t) & =  h_{+}(\boldsymbol{\theta},t) - \mathrm{i}\, h_{\times}(\boldsymbol{\theta},t) \label{eq:htotal}\\
        & =  \sum_{j=0}^{2}h_{j,+}(\boldsymbol{\theta},t) - \mathrm{i}\, h_{j,\times}(\boldsymbol{\theta},t),\label{eq:hsum} \\
        h_{j,+}(\boldsymbol{\theta},t) & = 
        H w_j \exp\left[-\frac{t}{T_j}\right] \cos \left(2\pi f_j t\left[1+\alpha_j t\right]+\psi_j \right). \label{eq:hmode}
\end{align}
        Here, ${\boldsymbol{\theta} = \{H,w_j,T_j,f_j,\alpha_j,\psi_j :  j \in [0,2] \}}$ are the model parameters where $H$ is the amplitude scaling factor and $w_j$ is the relative scaling factor for each mode, $j\in [0,2]$, such that $\sum_j w_j = 1$. 
        The initial frequency of each mode is given by $f_j$, $T_j$ are the damping times, $\psi_j$ are the initial phases, and $\alpha_j$ are the frequency drift terms. 
        The time, $t$, is defined such that $t=0$ corresponds to the coalescence time when the maximum of $h_+^2(\boldsymbol{\theta},t)+h_\times^2(\boldsymbol{\theta},t)$ occurs~\cite[e.g.][]{Read2013,Takami2015,Rezzolla2016,Easter2019}.
        The cross polarisation of the $j$th mode is generated by a $\pi/2$ phase shift on $h_{j,+}(\boldsymbol{\theta},t)$.
        Setting $\alpha_j=0$ allows detection of signals corresponding to the cross-polarisation model in \cite{Bauswein2016}. 
        These equations are a subset of the plus polarisation model in \cite{Bose2018} with the quadratic drift term set to zero and no explicit modulation of spectral peaks.
        \par
        
        We use nine post-merger numerical-relativity simulations from \cite{Dietrich2018} (see Appendix~\ref{appendix:a} for details), selecting only simulations with equal-mass progenitors where a nascent neutron star
        survives for at least $\sim\unit[25]{ms}$.
        For equal-mass systems, the tidal parameter of the neutron stars is related to the dimensionless compactness, $C=GM/(Rc^2)$, and the second Love number, $k_2$, as follows:
        \begin{align}
            \tilde{\Lambda} = \frac{2}{3} k_2 C^{-5}, \\
            \kappa_2^{\mathrm{T}} = \frac {1}{8}k_2 C^{-5},
        \end{align}
        where $\tilde{\Lambda}$ is the quadrupolar tidal deformability and $\kappa_2^{\mathrm{T}}$ is the total quadrupolar tidal coupling constant. 
        Here, $M$ is the neutron star mass, $R$ is the neutron star radius, $G$ is the gravitational constant, and $c$ is the speed of light. 
        The tidal properties of the progenitors can be estimated from the dominant post-merger frequency using relations found from numerical-relativity simulations with baryonic equations of state~\cite{Takami2015,Bauswein2019} (although see~\cite{Most2018b,Most2018,Bauswein2019} for the consequences of a phase transition to strange matter). 
        We discuss this more in Section~\ref{sec:ParameterEstimation}.      \par
        
        We inject numerical-relativity waveforms at various post-merger signal-to-noise ratios into a three-detector network (LIGO Hanford, Livingston, and Virgo) at design sensitivity for each interferometer~\cite{PSD:aLIGO,PSD:aVirgo}. 
        We inject the post-merger signal at a fixed time and fixed sky position, assuming that we know the coalescence time from the inspiral stage. 
        In Section~\ref{sec:sensitivity} we test this assumption by determining the uncertainty in the coalescence time for various signal-to-noise ratios.
        We use the {\sc{Bilby}} package  \cite{Ashton2019} with the {\sc{Dynesty}} sampler~\cite{Speagle2019} to sample posteriors, $p(\boldsymbol{\theta}|d)$, of the model parameters  using the likelihood, $\mathcal{L}(d | \boldsymbol{\theta})$, as follows:
        \begin{align}
        	p(\boldsymbol{\theta}|d) & =\dfrac{\mathcal{L}(d | \boldsymbol{\theta})\pi(\boldsymbol{\theta})}{\mathcal{Z}},\label{eq:posterior}\\
        	\mathcal{Z} & =\int_{\boldsymbol{\theta}} d\boldsymbol{\theta} \mathcal{L}(d | \boldsymbol{\theta})\pi(\boldsymbol{\theta}),\label{eq:evidence}\\
        	\mathcal{L}(d | \boldsymbol{\theta}) & \propto \exp\Big[-\Big<d(t)-h(\boldsymbol{\theta},t),d(t)-h(\boldsymbol{\theta},t)\Big>\Big]\label{eq:likelihood}. 
        \end{align}        
        Here, $d(t)=s(t)+n(t)$ is the numerical-relativity waveform, $s(t)$, injected into noise, $n(t)$. 
        We simulate ten different Gaussian noise realisations with {\sc{Bilby}}, to examine the response of the model to variations in detector noise. 
        We limit this to ten noise realisations to keep the computation time manageable.
        The priors on the model parameters are $\pi(\boldsymbol{\theta})$. 
        The noise-weighted inner product in Eq.~\ref{eq:likelihood} is defined by:
        \begin{align}
        	\left<h_1,h_2\right>\equiv4\, \mathrm{Re} \int df\frac{\tilde{h}_1(f)\tilde{h}^{*}_2(f)}{S_h(f)},\label{eq:inner_product}
        \end{align}
    where $S_h$ is the detector's noise power spectral density.
    We use a sampling frequency of 8192\,Hz to eliminate aliasing of the upper sidebands. 
    We use constrained priors to sort the maximum amplitude for $\tilde{h}_{j,+}(\boldsymbol{\theta},f)$, such that  $|\tilde{h}_{j,+}(\boldsymbol{\theta},f)|_{max} > |\tilde{h}_{j+1,+}(\boldsymbol{\theta},f)|_{max}$.
    This ensures that the mode zero ($j=0$) exponentially damped sinusoid corresponds to the dominant post-merger frequency. 
    Full details on the priors are given in Appendix~\ref{appendix:b}. 
    The optimal post-merger signal-to-noise ratio, $\rho_{\mathrm{opt}}$, is calculated from the quadrature sum of the optimal post-merger signal-to-noise ratio for each of the three detectors, $\rho_{\mathrm{opt},i}$ as follows:
\begin{equation}
    \rho_{\mathrm{opt}}^2 = \sum_{i \in  \mathrm{HLV}} \rho_{\mathrm{opt},i}^2\quad,\label{eq:OptimalSNR}
\end{equation}
    for $t \geq 0$.
    The matched filter signal-to-noise ratio for a single detector is given by:
\begin{equation}
    \rho_{mf} = \frac{\left<d, h(\boldsymbol{\theta})\right>}{\left< h(\boldsymbol{\theta}),h(\boldsymbol{\theta})\right>^\frac{1}{2}}\quad.\label{eq:MatchedFilterSNR}
\end{equation} \par

\section{Model Validation}\label{sec:modelfit}
     Figure~\ref{fig:TimeFrequencyResponse} shows the posterior waveforms in the time and frequency domain for the plus polarisation of two numerical-relativity post-merger simulations. The two gravitational-wave simulations, SLy-M1.350-$\Lambda$390 (THC:0036:R03, top) and LS220-M1.350-$\Lambda$684 (THC:0019:R05, bottom) are injected at a post-merger signal-to-noise ratio of 50. 
     These waveforms are chosen for compatibility with the inferred properties of $\Lambda$ from GW170817~\cite{GW170817Detection,Annala2018,Radice2018,Most2018,De2018,GW170817Properties}. 
     SLy-M1.350-$\Lambda$390 is a simulation of equal progenitor mass $1.35\,M_\odot$ neutron stars with tidal deformability, $\tilde{\Lambda}=390.1$ ($\kappa_2^{\mathrm{T}}=73.14$) and SLy equation of state. Similarly, LS220-M1.350-$\Lambda$684 has masses of $1.35\,M_\odot$, $\tilde{\Lambda}=683.8$ ($\kappa_2^{\mathrm{T}}=128.2$) and LS220 equation of state. \par
\begin{figure}[H]
        \centering
        \includegraphics[scale=0.5]{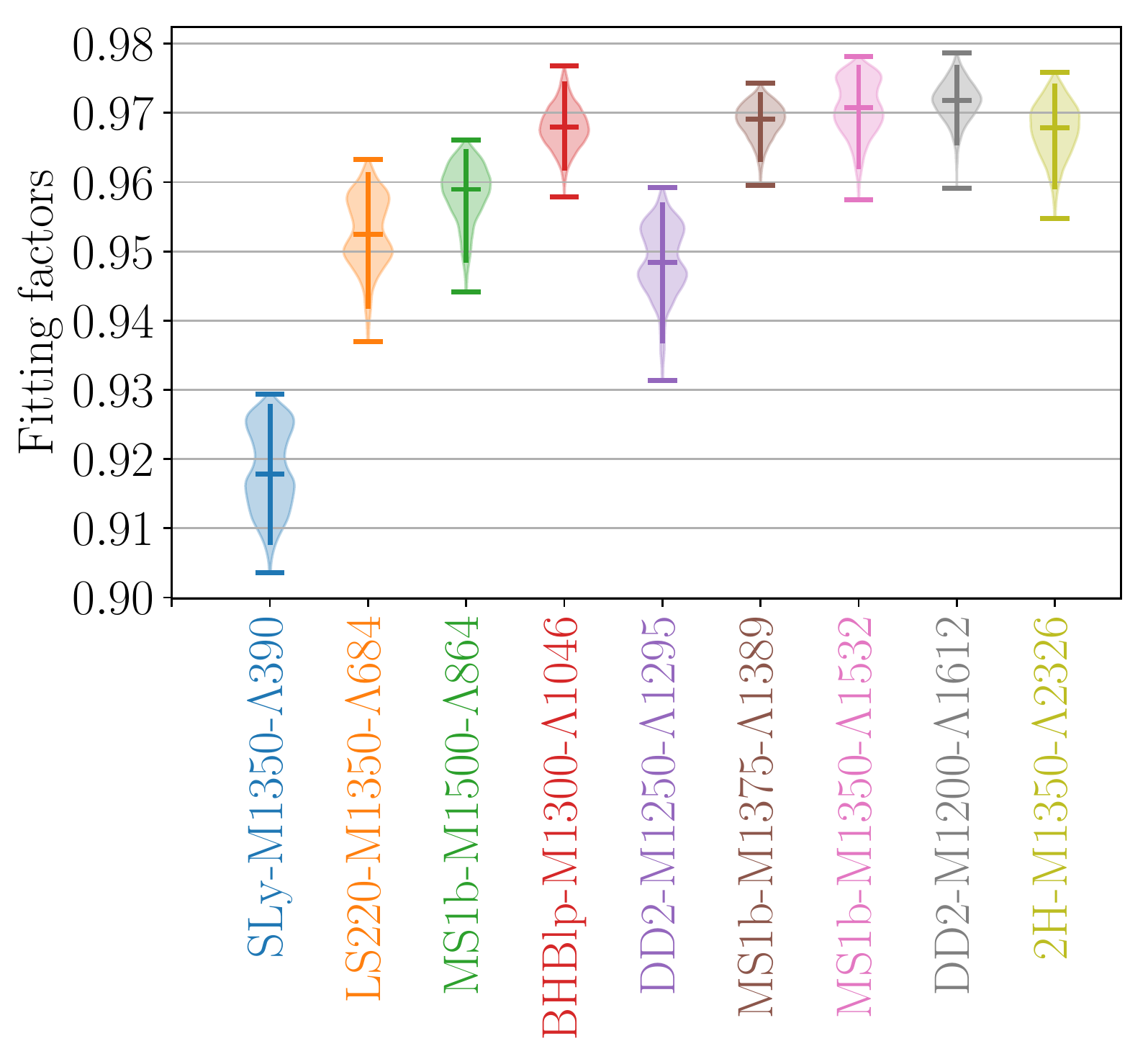}
        \caption{Fitting-factor distributions, $\mathcal{F}(d(t),h(\boldsymbol{\theta},t))$, for each post-merger numerical-relativity waveform. 
        The signal-to-noise ratio for the post-merger gravitational-wave strain for each waveform is 50. 
        The upper and lower horizontal bars represent 99.7\% confidence intervals. 
        The central horizontal bar shows the median value. 
        The thick vertical line shows the 95\% confidence intervals. 
        The median fitting factors range between 0.92 to 0.97 which corresponds to a reduction in detection rate from $22\%$ down to $9\%$ due to mismatch with the numerical-relativity injections.} 
        \label{fig:FittingFactors}
    \end{figure}     
     We generate posterior waveforms by randomly drawing samples from the posterior distribution $p(\boldsymbol{\theta}|d)$. The posterior waveforms are shown as  blue (top, SLy-M1.350-$\Lambda$390) and orange (bottom, LS220-M1.350-$\Lambda$684)  curves in Fig.~\ref{fig:TimeFrequencyResponse}. 
     The solid black curves show the injected numerical-relativity waveforms. 
     As can be seen in the time-response plots (Fig.~\ref{fig:TimeFrequencyResponse}, left), the posterior samples are tightly clustered around the numerical-relativity simulations, particularly for the first $\sim\,15\,$ms. 
     However, the phase of waveform SLy-M1.350-$\Lambda$390 is lost after $\sim\,15\,$ms (Fig.~\ref{fig:TimeFrequencyResponse}, upper left).
     \par 
     The frequency-response plots  are shown on the right side of Fig.~\ref{fig:TimeFrequencyResponse}, along with the amplitude spectral density of Advanced LIGO (dashed black curve) and Advanced Virgo (dotted black curve) at design sensitivity. 
     The primary frequency peaks are well recovered for both reference waveforms. 
     Two low frequency peaks of SLy-M1.350-$\Lambda$390 are resolved in preference to the upper frequency peak, whereas only one low frequency peak is resolved for LS220-M1.350-$\Lambda$684. The other two modes are located at the main frequency peak of LS220-M1.350-$\Lambda$684.
     \par
     To measure the extent of the waveform mismatch, we calculate the noise-weighted fitting factor between the injected numerical-relativity waveform, $d(t)$, and the posterior waveform, $h(\boldsymbol{\theta},t)$, ~\cite{apostolatos95}:
    \begin{equation}
    	\mathcal{F}(d(t),h(\boldsymbol{\theta},t))\equiv\frac{\left<d(t)|h(\boldsymbol{\theta},t)\right>}{\sqrt{\left<d(t)|d(t)\right>\left<h(\boldsymbol{\theta},t)|h(\boldsymbol{\theta},t)\right>}}.\label{eq:FF}
    \end{equation}
    The fitting factor, calculated with noise from one detector at Advanced LIGO design sensitivity~\cite{PSD:aLIGO}, quantifies the loss in signal-to-noise due to signal mismatch in relation to an optimal signal-to-noise ratio, Eq.~\ref{eq:OptimalSNR}. \par

    The median fitting factors are $0.92$ and $0.95$, for SLy-M1.350-$\Lambda$390 and LS220-M1.350-$\Lambda$684, respectively. As the detection rate scales as $\mathcal{F}^3$~\cite{apostolatos95}, the reduction in detection rate due to the above mismatch is 22\% and 14\% respectively for these two waveforms. \par
    
    The fitting factors for all nine numerical-relativity simulations are shown in Fig.~\ref{fig:FittingFactors}, with each simulation represented by a  different colour.
    Ten different Gaussian noise realisations are used for each numerical-relativity simulation.
    The 99.7\% confidence intervals for the fitting factors are shown by the upper and lower horizontal bars.
    The median value is shown by the central horizontal bar, and 95\% confidence intervals are indicated by thick vertical bars.
    Finally, the distribution of the fitting factors are shown by the width of the shaded areas.
    The lowest fitting factors, for simulation, SLy-M1.350-$\Lambda$390, have an average match of 0.92. 
    Other numerical-relativity injections have fitting factors of $0.95-0.97$. 
    The injection with the softest equation of state under-performs the other injections. 
    This is due to complex dynamics of the nascent neutron star in the first $\sim 2\,$ms.
\par

\section{Sensitivity}\label{sec:sensitivity}
\begin{figure}[H]
         \centering
         \includegraphics[scale=0.5]{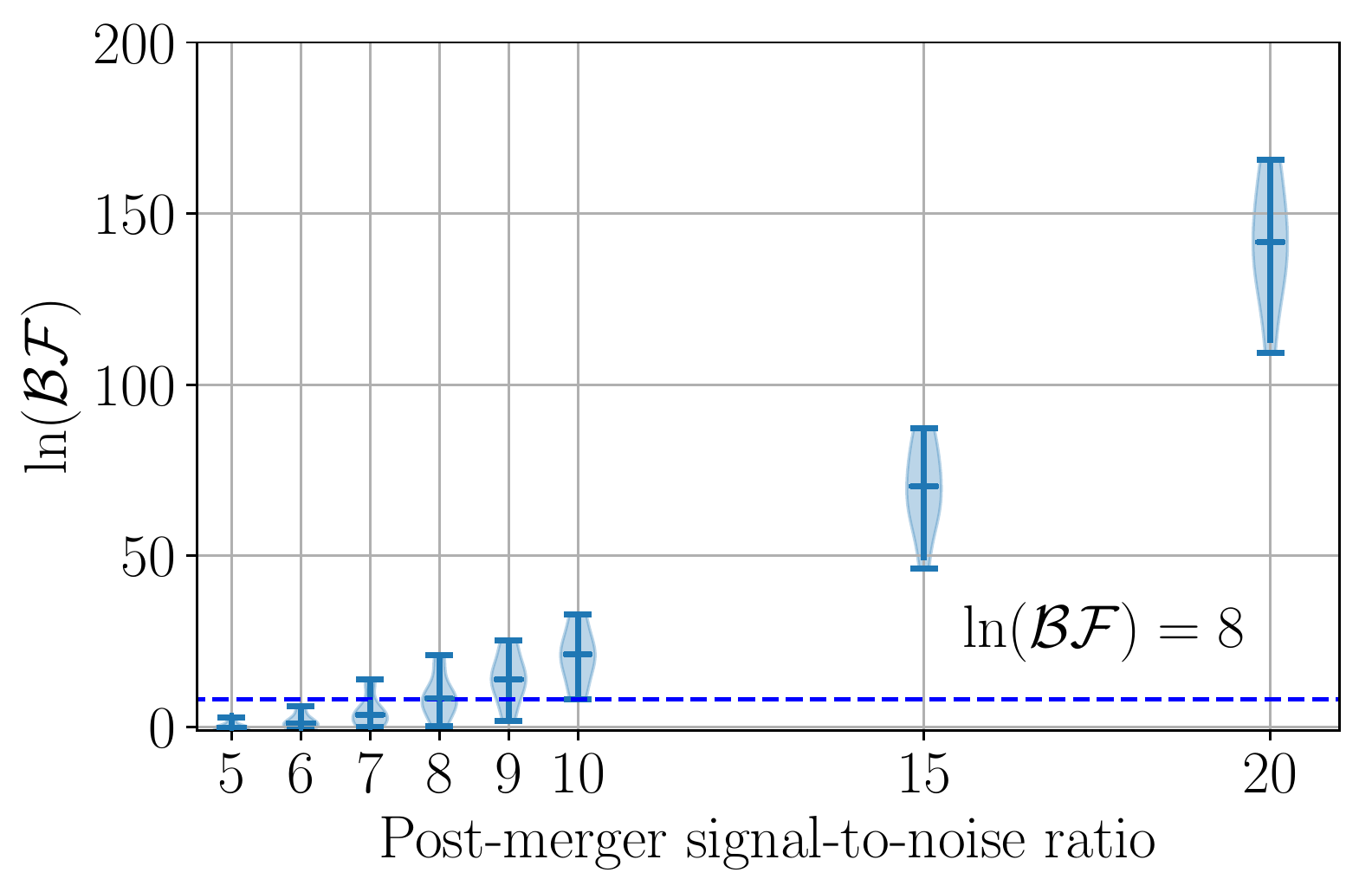}
         \caption{Natural logarithm of the Bayes factor comparing a signal hypothesis against a noise hypothesis plotted against the post-merger signal-to-noise ratio.
         The numerical-relativity waveform, SLy-M1.350-$\Lambda$390, is injected into ten different noise realisations at the specified signal-to-noise ratio.  
         The upper and lower horizontal bars show the 99.7\% confidence intervals of the log Bayes factor and the central horizontal bar shows the median value. 
         A post-merger signal-to-noise value of $\gtrsim 10$ is required to ensure strong evidence for a signal hypothesis ($\ln(\mathcal{BF})>8$). 
         Depending on the specific noise realisation there is some chance for strong signal evidence for post-merger signal-to-noise ratios as low as 7.}
         \label{fig:BFvsSNR}
     \end{figure}

    We calculate the Bayes factor between the signal hypothesis and a noise hypothesis to evaluate the sensitivity of our model. 
    We do this by injecting  the post-merger signal SLy-M1.350-$\Lambda$390 into ten different noise realisations at various signal-to-noise ratios. 
    The results are shown in Fig.~\ref{fig:BFvsSNR}.
    The distribution of the natural logarithm of the Bayes factor, $\ln(\mathcal{BF})$, is shown for each post-merger signal-to-noise ratio along with the 99.7\% confidence intervals (upper and lower horizontal bars) and the median value (middle horizontal bar). 
    We define that strong evidence for a signal hypothesis over a noise hypothesis corresponds to a Bayes factor exceeding 3000 ($\ln(\mathcal{BF})> 8.0$)~\cite[e.g.][]{Jeffreys61}. 
    In this case a signal hypothesis is 3000 times more likely than a noise hypothesis.
    This occurs with post-merger signal-to-noise ratios of $\gtrsim 10$.
    However, strong evidence for a signal can be obtained for post-merger signal-to-noise ratios of $\approx 7\Hyphdash*9$, depending on the specific noise realisation. \par
\begin{figure}[H]
         \centering
         \includegraphics[scale=0.5]{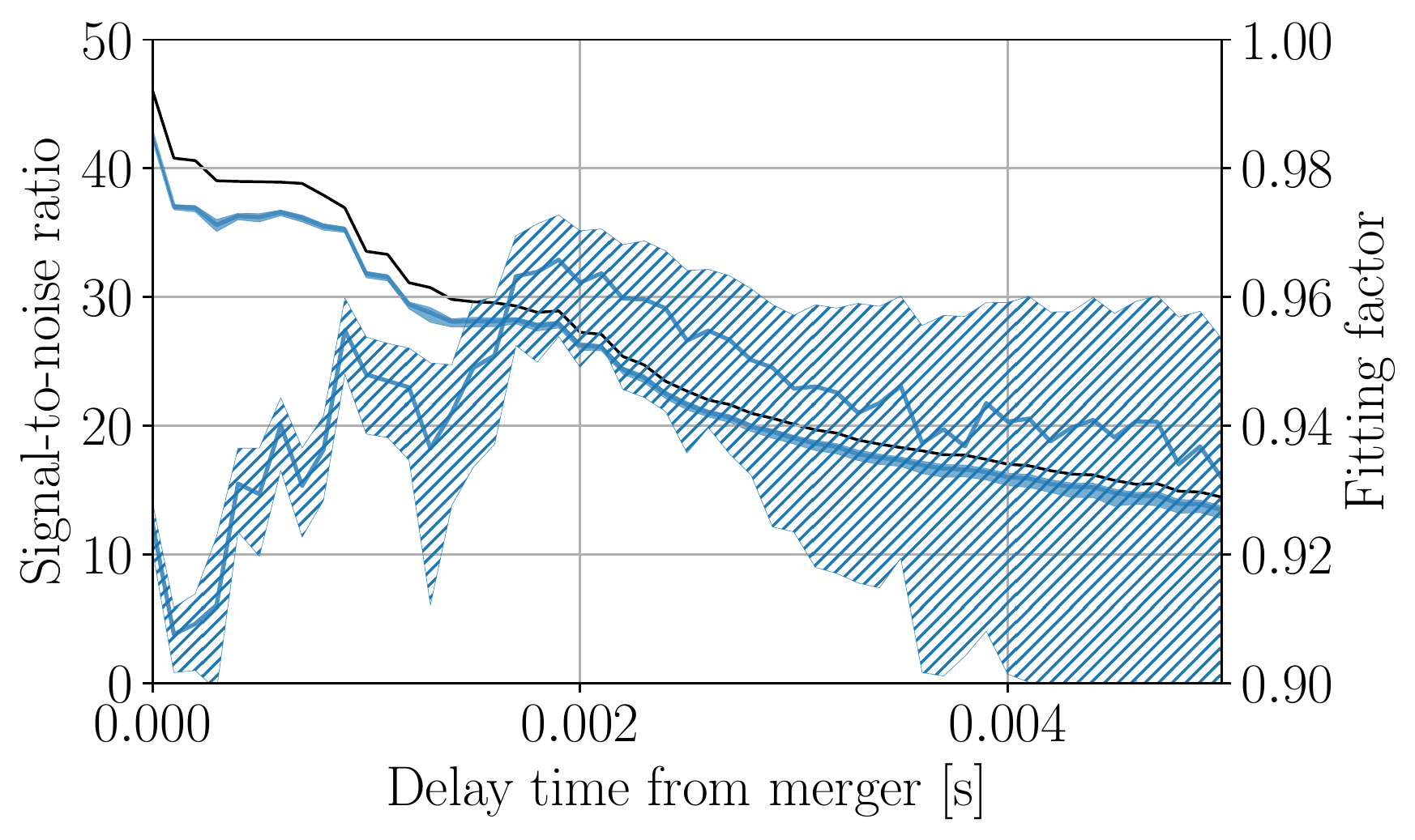}
         \caption{Variation of fitting factor (hatched blue) and matched filter  signal-to-noise ratio (solid blue) for differing values of the delay time after the time of coalescence. The shaded regions show 95\% confidence intervals. The optimal signal-to-noise ratio is also shown (solid black). Although the fitting factor is lower when the entire post-merger signal is used, the matched filter signal-to-noise ratio is largest. The fitting factor is lower for smaller delay-times due to the complex dynamics of the nascent neutron-star.}
     \label{fig:OF_MF_SNR_FF5ms}
     \end{figure}

\begin{figure}[H]
         \centering
         \includegraphics[scale=0.5]{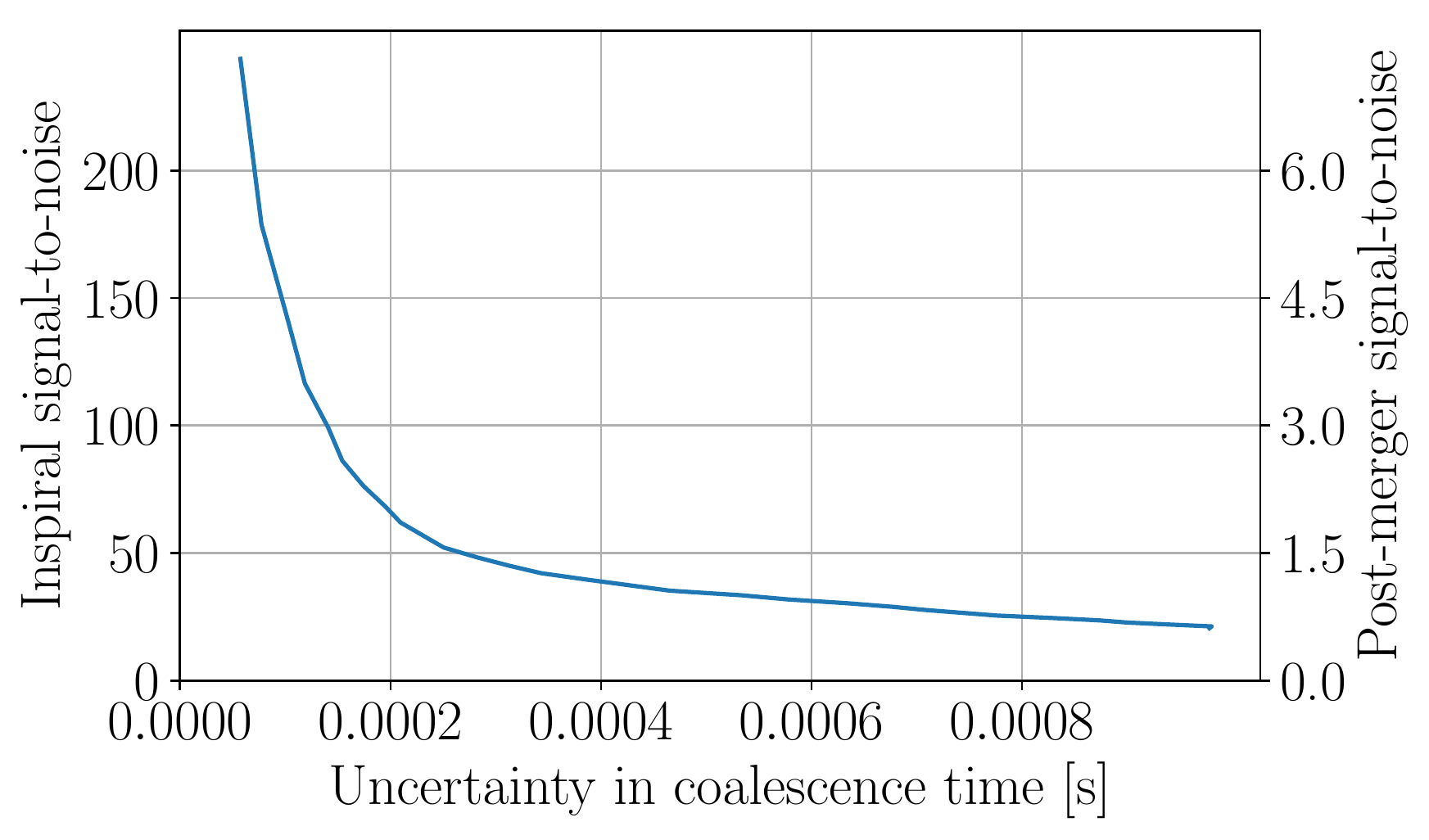}
         \caption{Uncertainty in coalescence time plotted with signal-to-noise ratios. The coalescence time uncertainty has been determined by a Fisher matrix approximation. The left axis is the signal-to-noise ratio for a two detector network of Advanced LIGO at design sensitivity for a binary neutron star inspiral. The right axis is the post-merger signal-to-noise ratio for a three detector network of Advanced LIGO and Advanced Virgo at design sensitivity using numerical-relativity simulation SLy-M1.350-$\Lambda$390. Post-merger signal to noise ratios above 6.0 have coalescence time uncertainties of less than 0.1\,ms.}
         \label{fig:TCOALESCENCE}
     \end{figure}

     An important consideration for our signal model is the uncertainty in the coalescence time as measured from the gravitational-wave inspiral signal. 
     This determines how close we can get to the true coalescence time for the binary neutron star merger.
     In Fig.~\ref{fig:OF_MF_SNR_FF5ms} we investigate the model performance to uncertainties in the coalescence time.
     We show how the fitting-factor and matched-filter signal-to-noise ratio change when starting the adopted model at various times after the coalescence time.
     We multiply the numerical-relativity injection, $d(t)$, by the Heaviside step function, $\mathcal{H}(t-t_{delay})$, and evaluate the model, $h(\boldsymbol{\theta},t-t_{delay})$, for $t \geq t_{delay}$.
     The matched filter signal-to-noise ratio is calculated using Eq.~\ref{eq:MatchedFilterSNR} with a single detector at Advanced LIGO sensitivity.
     We use numerical-relativity injection, SLy-M1.350-$\Lambda$390, selected due to compatibility with the tidal parameters inferred from GW170817.
     A delay time of zero includes the entire post-merger waveform, whereas a delay time of $2\,$ms excludes the first $2\,$ms of the injection after the coalescence time. 
     The fitting-factor is lower ($\sim 0.91$) for small delay times and increases to $\sim\,0.96$ at $2\,$ms. 
     The fitting factor is lower in the first $2\,$ms due to complex dynamics of the nascent neutron star.
     In Fig.~\ref{fig:OF_MF_SNR_FF5ms}, the matched-filter signal-to-noise ratio is almost monotonically decreasing as expected. 
     Even though the fitting-factors are lower at zero delay time, the matched-filter signal-to-noise ratio is at maximum. 
     Therefore, from a sensitivity perspective, a minimum delay time is preferred.
     To estimate the uncertainties of the time of coalescence of the inspiral signal as a function of the signal-to-noise ratio of the inspiral signal, we use a Fisher matrix approximation. 
     We assume that the signal parameters $\boldsymbol{\vartheta}$ follow a Gaussian distribution: 
\begin{equation}
    p \left( \Delta \boldsymbol{\vartheta} \right) \propto \exp \left[-\frac{1}{2}\Gamma_{ij}\Delta \vartheta^i \Delta \vartheta^j\right].
\end{equation}
    Here, $\Delta \vartheta_i = \vartheta^i-\hat{\vartheta}^i$, $\hat{\vartheta}^i$ are the best fit inspiral parameters and $\Gamma_{ij} = (\partial h/ \partial \vartheta_i | \partial h/ \partial \vartheta_j)$ is the expected Fisher information matrix. 
    The estimated errors of the parameters, $\vartheta_i$, are obtained by taking the diagonal elements of the Fisher information matrix. 
    The relevant parameters within our approximation are $\boldsymbol{\vartheta} = (\mathcal{M}, q, \phi_c, \tilde{\Lambda}, t_c,H)$, where $\mathcal{M}$ is the chirp mass, $q$ is the mass ratio, $\phi_c$ is the phase of coalescence.
    The average-weighted tidal deformabilty is  $\tilde{\Lambda}$, $t_c$ is the time of coalescence and $H$ is the amplitude of the inspiral waveform.
    We calculate the errors on $\vartheta_i$ assuming an equal mass 1.4\,$M_\odot$ non-rotating progenitor system. 
    The expected uncertainties for the coalescence time are shown in  Fig.~\ref{fig:TCOALESCENCE}. 
    The left axis shows the inspiral signal-to-noise ratio for an optimally oriented source into a two detector LIGO network at design sensitivity.
    We use Fig.~8 from \cite{Martynov2019} to determine the luminosity distance, $D_L$, from the inspiral signal-to-noise ratio. 
    We calculate the product of $D_{L0}\approx 475$\,Mpc (at $z\approx 0.1$) with the corresponding inspiral signal-to-noise ratio, $\rho_{inspiral,0}\approx 7$.
    We inject the numerical-relativity post-merger waveform, SLy-M1.350-$\Lambda$390, at luminosity distance, $D_L = D_{L0}\left(\rho_{inspiral,0}/\rho_{inspiral}\right)$, and evaluate the post-merger signal-to-noise ratio using the Advanced LIGO and Virgo detector network at design sensitivity.
    The right axis in  Fig.~\ref{fig:TCOALESCENCE} shows the corresponding post-merger signal-to-noise ratio.\par
    For post-merger signal-to-noise ratios larger than 6, the uncertainty in the coalescence time is less than 0.1\,ms.
    This shows that, for post-merger signal-to-noise ratios of interest in this work, the coalescence time is similarly constrained. 
    The uncertainty in coalescence time can be related to Fig.~\ref{fig:OF_MF_SNR_FF5ms} to show that the resultant matched filter signal-to-noise ratio is not significantly reduced due to the uncertainty in the coalescence time.

\section{Parameter  estimation}\label{sec:ParameterEstimation}
\begin{figure}[H]
         \centering
         \includegraphics[scale=0.5]{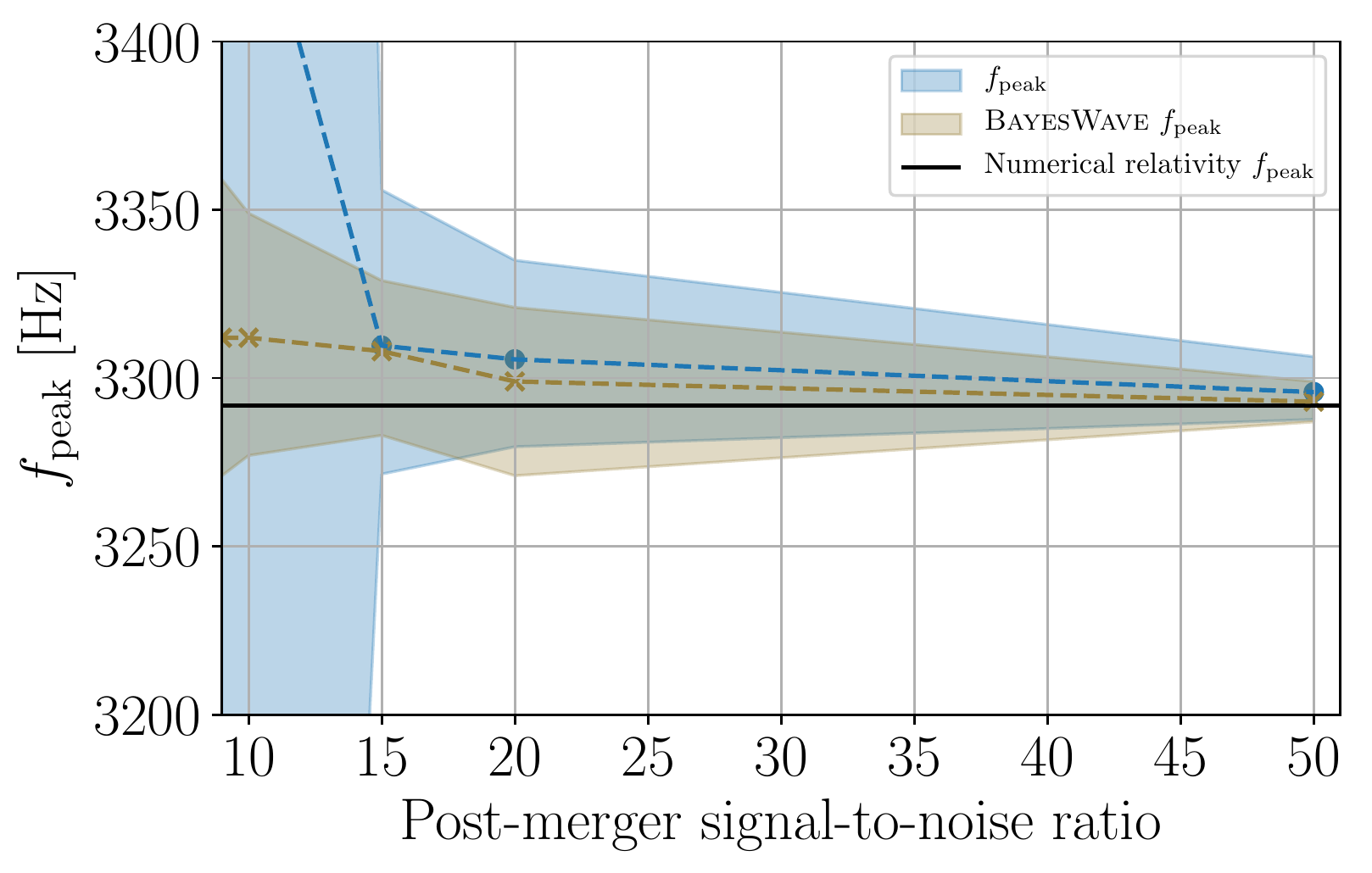}
         \caption{Primary post-merger frequency comparison between our model (blue) and \textsc{BayesWave} (brown). The posteriors are plotted against the post-merger signal-to-noise ratio for injection SLy-M1.350-$\Lambda$390. The 95\% confidence intervals are shaded. The median points are shown as blue dots and brown crosses, for our model and \textsc{BayesWave}, respectively. The frequency corresponding to the peak of the spectral response of the injection is also shown (solid black line).
         \label{fig:BWFREQvsSNR}}
     \end{figure}

    Estimation of the primary post-merger frequency is another important indicator of the model performance.
    We estimate this by calculating posteriors of the peak frequency, $f_{\mathrm{peak}}$, of the dominant mode.
    Fig.~\ref{fig:BWFREQvsSNR} shows posteriors of $f_{\mathrm{peak}}$ as a function of post-merger signal-to-noise ratio.
    These have been calculated for an injection of SLy-M1.350-$\Lambda$390 at post-merger signal-to-noise ratios  of $\ge 9$.
    The noise realisation was kept the same for all injections. 
    Blue shading indicates regions of 95\% confidence intervals and the median values are shown as blue dots.
    The frequency corresponding to the maximum value of the characteristic strain spectrum of the  numerical-relativity signal, $|\tilde{s}_+(f)|\sqrt{f}$, is shown as a black horizontal line.
    This can be thought of as an approximation of the true injected value of $f_{\mathrm{peak}}$.
    The $f_{\mathrm{peak}}$ frequency is constrained within 95\% confidence intervals to $3310\pm^{46}_{38}$\,Hz at a post-merger signal-to-noise ratio of 15 which corresponds to $\pm^{1.4}_{1.2}\%$.
    At a post-merger signal-to-noise ratio of 50, the precision increases to $3296\pm^{11}_{8}$\,Hz $(\pm^{0.3}_{0.2}\%)$.
    The posteriors for $f_0, \alpha_0, f_1$ and $\alpha_1$, determined for all numerical-relativity injections at a post-merger signal-to-noise ratio of 50, are shown in Figs.~\ref{fig:Corner1}-\ref{fig:Corner9} in Appendix~\ref{appendix:c}.\par

    We also analyse injections of SLy-M1.350-$\Lambda$390 using \textsc{BayesWave} \cite{Cornish2015,Littenberg2015}. \textsc{BayesWave} uses a variable number of Morlet-Gabor wavelets to model the signal, where both the number and the properties of the wavelets are marginalised over.
    This is an established method for post-merger studies~\cite{GW170817Properties,GW190425Detection}.
    References~\cite{Chatziioannou2017,Torres-Rivas2019} have performed simulations using \textsc{BayesWave} to infer the post-merger properties of binary neutron star mergers.
    We compute the posteriors of the spectral frequency peak, $f_{\mathrm{peak}}$, using \textsc{BayesWave} following~\cite{Chatziioannou2017}. Here, $f_{\mathrm{peak}}$, the frequency of the highest peak in the Fourier power spectrum of the signal, is determined for each sample from the \textsc{BayesWave} posterior. 
    For samples that do not have a peak, $f_{\mathrm{peak}}$ is computed using  random draws from its prior~\cite{Chatziioannou2017}.
    Figure~\ref{fig:BWFREQvsSNR} shows the 95\% confidence intervals of $f_{\mathrm{peak}}$ for each post-merger signal-to-noise ratio in brown. The median values are shown as brown crosses.
    The \textsc{BayesWave} frequency posteriors are consistent with~\cite{Chatziioannou2017,Torres-Rivas2019}.
    The posteriors for $f_{\mathrm{peak}}$ are similarly constrained for both \textsc{BayesWave} and our model for post-merger signal-to-noise ratios of $\gtrsim 20$. 
    \textsc{BayesWave} is more  constrained for post-merger signal-to-noise ratios of $\sim 9 - 15$.
    Both methods are able to recover the injected dominant post-merger frequency.
    \textsc{BayesWave} can generate very high fitting factors; the fitting factors for SLy-M1.350-$\Lambda390$ at a post-merger signal-to-noise ratio of 50 are $\approx 0.99$.
    The dimensionality of \textsc{BayesWave} is $\sim 90\  (\sim 18$ wavelets) at this post-merger signal-to-noise ratio.
    The dimensionality of our adopted model is 15 with fitting factors of $\approx 0.92$ for SLy-M1.350-$\Lambda390$.
    Furthermore, \textsc{BayesWave} can generalise to any signal (e.g. glitches).
    In contrast, our model has been developed to suit a post-merger gravitational-wave signal.
    The parameters in our model are interpretable: 
    for example, in Fig.~\ref{fig:Corner1}, the $\alpha_0$ value for SLy-M1.350-$\Lambda390$ is $-1.60\pm^{0.50}_{0.26}$ which
     shows that the frequency of the dominant gravitational-wave mode is decreasing.
\begin{figure}[H]
         \centering
         \includegraphics[scale=0.5    ]{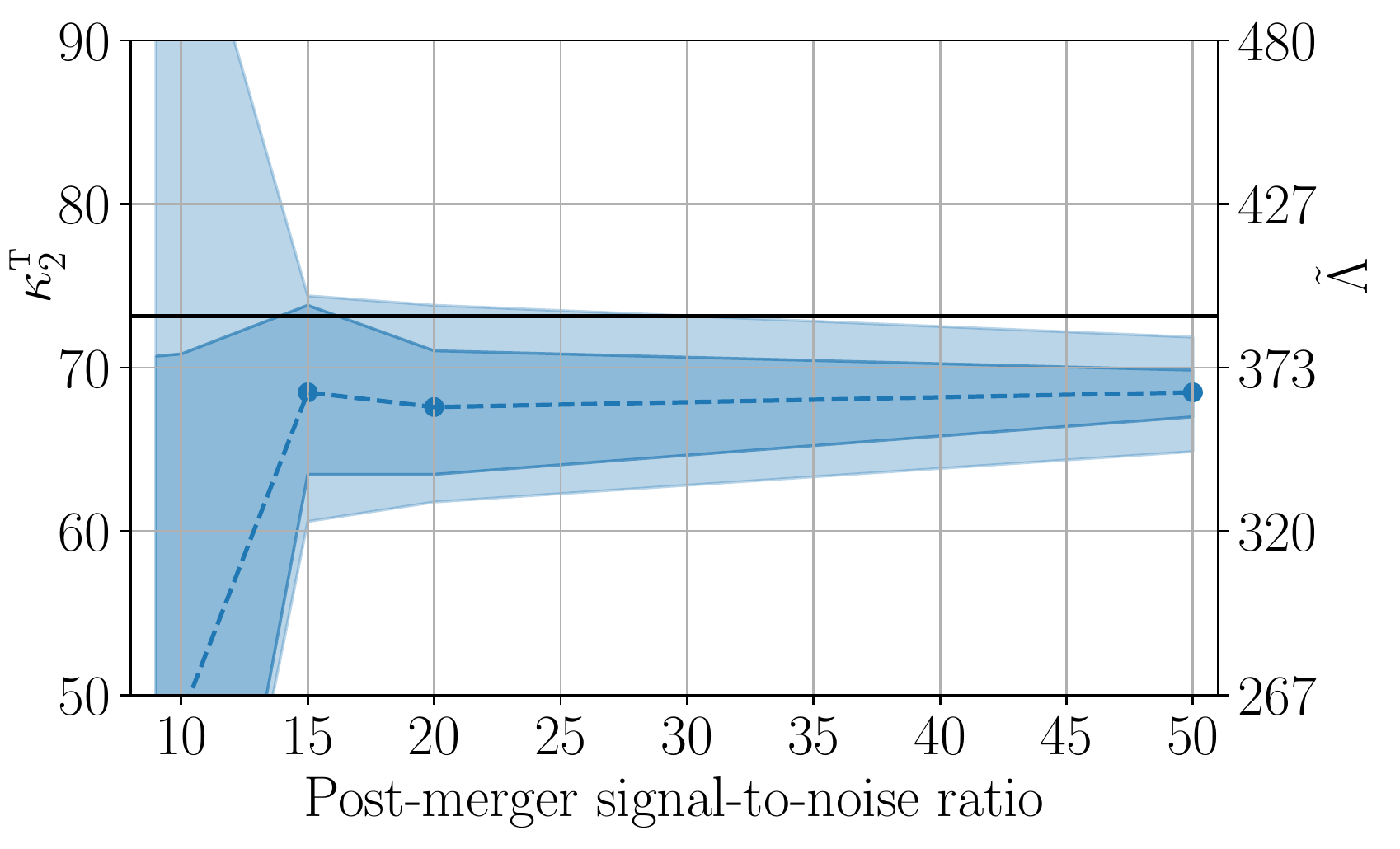}
         \caption{Tidal coupling constant posteriors versus post-merger signal-to-noise ratio for numerical-relativity waveform SLy-M1.350-$\Lambda$390. The tidal coupling constant is inferred from the hierarchical model~\cite{Easter2019} using the magnitude of the posterior waveforms, $|h_+(\boldsymbol{\theta},t)|$. The 68\% (dark blue) and 95\% (light blue) confidence intervals are shown along with the median values (blue dots). The true value for $\kappa_2^\textsc{t}$ is shown as the solid horizontal line. 
         The corresponding tidal deformability values are shown on the secondary vertical axis.}
         \label{fig:KappavsSNR}
     \end{figure} 
    The hierarchical model from~\cite{Easter2019} allows a bidirectional relationship between equal mass progenitor neutron star properties $(C,M,\kappa_2^\textsc{t})$ and numerical-relativity post-merger simulations. 
    This is achieved by a two step process. 
    Firstly, the progenitor properties are used to solve $\bar{C}(M,\kappa_2^\textsc{t})=C$ using a power-law relationship.  
    Secondly, the model parameters, $\boldsymbol{\Theta}$, are determined by solving $h_c = \boldsymbol{\Theta}\, \boldsymbol{X}(\bar{C}(M,\kappa_2^\textsc{t}),M,\kappa_2^\textsc{t})$, where $h_c$ is the numerical-relativity amplitude spectra for the characteristic strain $(h_c(f)=|\tilde{h}(f)|\sqrt{f})$.  
    Here, $\boldsymbol{X}(\bar{C} (M,\kappa_2^\textsc{t}),M,\kappa_2^\textsc{t})$ is a design matrix derived from the progenitor properties $M$ and $\kappa_2^\textsc{t}$.\par

    We use the posteriors from Section~\ref{sec:modelfit} to calculate the amplitude of the characteristic spectrum $|\tilde{h}_{+}(\boldsymbol{\theta},f)|\sqrt{f}$ and use the trained model, $\boldsymbol{\Theta}$,  to determine the hierarchical model posteriors on $\kappa_2^\textsc{t}$ and $C$.
    The cross-polarisation waveforms are discarded because the hierarchical model only uses the magnitude of the spectra, and  $|\tilde{h}_{+}(\boldsymbol{\theta},f)|=|\tilde{h}_{\times}(\boldsymbol{\theta},f)|$.
    The hierarchical model, $\boldsymbol{\Theta}$, was previously trained on 35 numerical-relativity simulations from~\cite{Rezzolla2016}, a distinct set of numerical-relativity simulations to those used in this paper.
    Therefore, this is an out-of-sample model validation. \par
  \begin{figure}[H]
         \centering
         \includegraphics[scale=0.5]{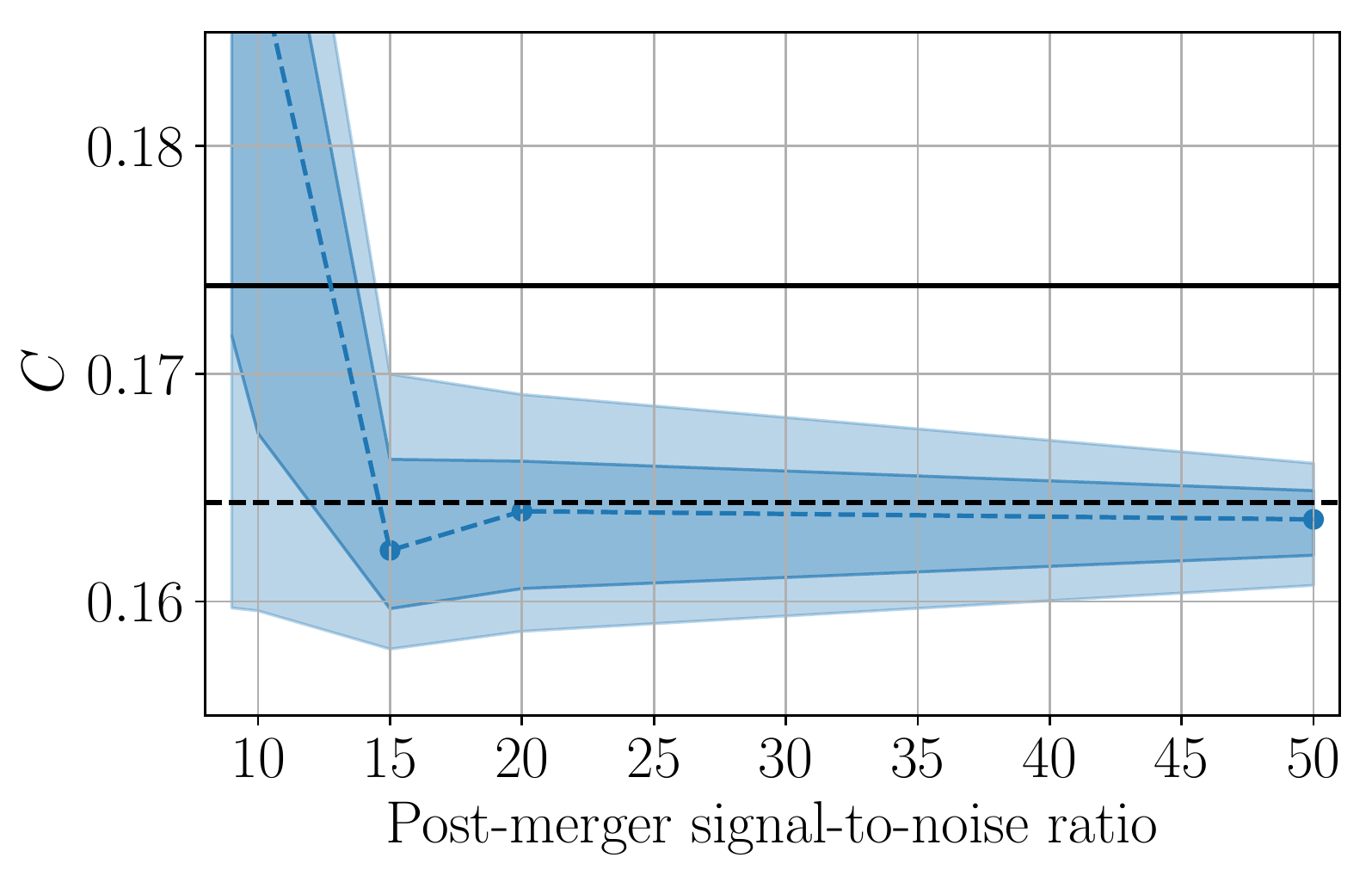}
         \caption{Compactness posteriors versus post-merger signal-to-noise ratio for numerical-relativity waveform SLy-M1.350-$\Lambda$390. The 68\% (dark blue) and 95\% (light blue) confidence regions are shown. The compactness is inferred from the hierarchical model~\cite{Easter2019}. The compactness inferred directly from the numerical-relativity waveform (dashed black line) and the compactness of the progenitor neutron stars (solid black line) are also shown. See the text for an explanation of the offset in these two values.}
         \label{fig:CompactnessvsSNR}
     \end{figure}   
   Figure~\ref{fig:KappavsSNR} shows the inferred posteriors for $\kappa_2^\textsc{t}$ with 68\% and 95\% confidence intervals in dark blue and light blue respectively. 
    The true injected value of $\kappa_2^\textsc{t}$, is shown by the horizontal solid black line and the median values as blue dots. 
    The  vertical axis shows both the quadrupolar tidal coupling constant (left axis) and the quadrupolar tidal deformability (right axis). 
    % The value of $\kappa_2^\textsc{t}$ inferred directly by the hierarchical model from the numerical-relativity simulation is shown by the horizontal black dotted line.
    The values inferred for the tidal parameters of the progenitor neutron stars are lower than the true value of the numerical-relativity injection, though the 95\% confidence interval excludes the true value only at a post-merger signal-to-noise ratio of 50.  
    The tidal coupling constant at a post-merger signal-to-noise ratio of 15 is  constrained at 95\% confidence intervals  to $68.5^{+5.9}_{-7.9}$, which tightens  to $68.5^{+3.4}_{-3.6}$ for  a post-merger signal-to-noise ratios of 50. \par
     There are a number of factors that will impact on the performance of the hierarchical model.
     Firstly, the numerical-relativity spectra from~\cite{Rezzolla2016}, which were used in~\cite{Easter2019} to train the model, are a distinct set of simulations to those in use in this paper~\cite{Dietrich2018}.
     Specifically, waveform SLy-M1.350-$\Lambda$390 is available in both sets of numerical-relativity simulations,~\cite{Dietrich2018} and \cite{Rezzolla2016}, and, although the primary post-merger peak occurs at the same frequency, the spectral response for the other frequencies are quite different.
     Secondly, the hierarchical model is an approximate model, and was only trained on 35 waveforms; a relatively small training set (for details, see Ref.~\cite{Easter2019}).
     Thirdly, the simulation outputs can be dependent on the spatial and temporal resolution, which can lead to waveform changes related to parameters like collapse time, primary oscillation frequency and decay time constants.\par

     Hierarchical model posteriors for the compactness, $C$, are shown in Fig.~\ref{fig:CompactnessvsSNR}. 
     The 68\% and 95\% confidence intervals are shaded dark blue and light blue  respectively, and the median values are shown with blue dots.
     The true value corresponding to the injected numerical-relativity simulation is shown as a horizontal solid black line.
     The value inferred from the numerical-relativity simulation using the hierarchical model is shown as a horizontal black dashed line. 
     The hierarchical posteriors for $C$ are clustered around the value inferred directly from the hierarchical model for the numerical-relativity simulation.
     In this case, the mismatch between the posteriors and the true value is more significant.
     The reasons for this are the same three reasons outlined above.
     The compactness has been constrained to $0.162^{+0.007}_{-0.004}$ at post-merger signal-to-noise ratios of 15 tightening to  $0.164^{+0.002}_{-0.003}$ at signal-to-noise ratios of 50 to 95\% confidence intervals. 
     The posteriors for the compactness, $C$, only narrow moderately as the post-merger signal-to-noise ratio is increased.
 
\section{Discussion}
    We use an analytical model to characterise gravitational-wave strain from nine numerical-relativity simulations selected such that the post-merger oscillations persist for $\sim 25$\,ms. 
    The median noise-weighted fitting factors for the posterior waveforms range between $0.92\,$\nobreakdash-$\,0.97$ for injections with post-merger signal-to-noise ratios of 50. 
    This corresponds to a loss in detection rate of  $22\,$\nobreakdash-$\,12\%$ when compared to a signal without mismatch.
    We measure the Bayes factor in favour of signal detection with numerical-relativity simulation SLy-M1.350-$\Lambda$390 and find that successful detections occur with post-merger signal-to-noise ratios of $\ge 10$ with possible detections as low as  post-merger signal-to-noise ratios of 7, depending on the specific noise realisation. 
    This indicates that this model could be used for parameter estimation and detection if a post-merger signal louder than signal-to-noise ratio of 10 was  coincident with an inspiral detection. 
    We find that this corresponds to a distance of $\sim\,10$\,Mpc for an optimally oriented system using a three-detector network (LIGO Hanford, Livingston, and Virgo) at design sensitivity.
    \par
    We determine that starting the model at the time of coalescence results in the maximum matched filter signal-to-noise ratio even though the fitting factors are lower in the vicinity of the merger due to the dynamics of the nascent neutron star.
    We find that the uncertainty in the time of coalescence for the inspiral of the progenitor neutron stars is less that 0.1\,ms for a post-merger signal-to-noise ratio of $\ge\,6$ and show that this corresponds to a maximum matched-filter signal-to-noise ratio. \par

    The gravitational-wave strain of the inspiral can constrain the equation of state for the cold neutron star at the high inspiral signal-to-noise ratios ($\gtrsim 200$) required for post-merger detection of the remnant (see Fig.~\ref{fig:TCOALESCENCE}). 
    This can place additional constraints on the priors for the dominant post-merger frequency. 
    However, a phase transition in the hot post-merger remnant~\cite{Most2018b,Most2018,Bauswein2019}, and uncertainty in the numerical-relativity calculations due to computational trade-offs, may result in a post-merger gravitational-wave signal that is quantitatively different than the  numerical-relativity simulations.
    With this in mind, we assume a more general, agnostic set of priors (see Appendix~\ref{appendix:b}).
    \par
 
    Using numerical-relativity waveform SLy-M1.350-$\Lambda$390, selected for its compatibility with $\Lambda$ values determined from GW170817~\cite[e.g.][]{GW170817Detection,Annala2018,Radice2018,Most2018,De2018,GW170817Properties}, we constrain the primary post-merger frequency to a range of $3310$\,Hz$\pm_{1.2}^{1.4}\%$ for 95\% confidence intervals at post-merger signal-to-noise ratios of 15.
    The precision increases to $3296$\,Hz$\pm_{0.2}^{0.3}\%$ for post-merger signal-to-noise ratios of 50. 
    We show that our model and \textsc{BayesWave} similarly constrain the dominant post-merger frequency, $f_{\mathrm{peak}}$, for post-merger signal-to-noise ratios of $\gtrsim 20$.
    For post-merger signal-to-noise ratio of $\sim 9 - 15$ \textsc{BayesWave} is better able to constrain $f_{\mathrm{peak}}$.
    We generate fitting factors of  $\approx 0.99$ using \textsc{BayesWave} for SLy-M1.350-$\Lambda$390 at a post-merger signal-to-noise ratio of 50.
    The corresponding fitting factors from our model are $\approx 0.92$.
    The dimensionality of the \textsc{BayesWave} posterior reconstruction is significantly larger than our analytic; $\sim 90$ dimensions for \textsc{BayesWave} cf. 15 for ours.
    Moreover, our adopted model is interpretable and can supply additional information about the individual modes (e.g. frequency drifts and exponential damping time constants).
     \par
    We use the hierarchical model from \cite{Easter2019}, which has been trained on numerical-relativity waveforms from ~\cite{Rezzolla2016}, to determine posteriors for $\kappa_2^\textsc{t}$ and $C$. 
    We obtain 95\% confidence intervals on $\kappa_2^\textsc{t}$ (and $\tilde{\Lambda}$) of $\pm^{9}_{12}\%$ at a post-merger signal-to-noise ratio of $15$ with increasing precision to $\pm\, 5\%$ at a post-merger signal-to-noise ratio of 50. 
    The 95\% confidence intervals on $C$ range from $\pm^{4.3}_{2.7}\%$ at post-merger signal-to-noise ratios of 15 to $\pm^{1.5}_{1.8}\%$ at post-merger signal-to-noise ratios of 50.
    However, due to a bias in the hierarchical model, the injected value for $C$ is outside the 95\% confidence interval. \par
    It should be noted that the inferred posteriors for $C$ are centred around the inferred values predicted by the hierarchical model.
    This indicates a bias in the hierarchical model that can be explained by three factors.
    Firstly, the numerical-relativity simulations are quite different between those used to train the hierarchical model~\cite{Rezzolla2016} and the waveform used for parameter estimation, SLy-M1.350-$\Lambda$390~\cite{Dietrich2017,Radice2016}. 
    The numerical-relativity simulations used to train the hierarchical model were homogeneous, changing only the equation of state and the progenitor masses between simulations, keeping other simulation parameters the same.
    Secondly, only 35 waveforms were used to train the hierarchical model which is a minimal training set.    
    Thirdly, waveforms generated from numerical-relativity simulations are dependent on resolution.
    Increasing the resolution can result in changes in both the time domain gravitational-wave strain as well as the corresponding spectral response (e.g. collapse time changes with resolution).
    It should also be emphasised that, because the numerical-relativity simulations are drawn from independent sources, the posteriors of $\kappa_2^\textsc{t} (\tilde{\Lambda})$ and $C$ are true out-of-sample estimates.
    We expect the estimates of $C$ and $\kappa_2^\textsc{t}$ to become more consistent with the injected value as the training set is increased in size and covers more system and progenitor properties. \par
    
    In addition to the aforementioned analytical models~\cite{Bauswein2016,Bose2018}, other work have generated analytical post-merger gravitational-wave  models. 
    In \cite{Hotokezaka2013}, a model  was generated for the time-based amplitude and phase of the complex gravitational-wave strain using a smooth piece-wise function for the amplitude. 
    The time-based phase was fit by the combination of a polynomial and exponentially-damped sinusoid using an iterative CMA-ES (covariance matrix adaption evolution search) fitting algorithm.
    The maximum fitting factors were  calculated in the time domain without noise weighting and are not directly comparable to the noise-weighted fitting factors calculated with Eq.~\ref{eq:inner_product}.
    Even so, the maximum fitting factors were  $\sim\!0.92\,$\nobreakdash-$\,0.98$ for 95\% of waveforms.\par
    
    A frequency-domain model was introduced in~\cite{Messenger2014} from analysing the major spectral peaks of the whitened power spectrum.
    The power of the dominant post-merger frequency peak was estimated by a trapezoidal structure and the model parameters were determined with a least-squares algorithm. 
    No fitting factors were calculated in this reference, as the goal was estimating source red-shifts.
    This model was extended in~\cite{Takami2015} to add a Gaussian component to the fundamental post-merger frequency using a nonlinear least-squares fit. 
    The goal of the fits in~\cite{Takami2015} were qualitative, rather than quantitative and no fitting factors were stated.\par
    
    The model used in~\cite{Bauswein2016} consists of three exponentially damped sinusoids centred at frequencies ($f_{2-0}, f_{\mathrm{spiral}}, f_{\mathrm{peak}})$ which are described in Section~\ref{sec:introduction}.
    In contrast, the model introduced in~\cite{Bose2018}, consists of two exponentially damped sinusoids, the first centred on $f_1$ which is modulated by frequency $f_{1e}$, and the second is centred on the dominant post-merger frequency, $f_2$, with a linear and quadratic frequency drift terms. 
    This model produced fits of $\sim 80\Hyphdash*94\%$. 
    In \cite{Tsang2019}, a frequency-domain model was developed for a single damped-sinusoid.
    This model was based on three or six parameters and used Bayesian inference to estimate the parameters. 
    They obtained fitting factors of $\sim\!0.60\,$\nobreakdash-$\,0.98$. 
    Reference~\cite{Breschi2019} parameterised the instantaneous amplitude and phase of the time-based gravitational-wave strain. 
    Their model uses a rational-polynomial fit based on the progenitor properties ($M_1$, $M_2$, $\kappa_2^\textsc{t}$) derived in~\cite{Bernuzzi2015,Zappa2018,Dietrich2019}.
    They achieved fitting factors of $\sim\!0.30\,$\nobreakdash-$\,0.85$ in zero noise.\par
    The fitting factors obtained in our paper compare favourably to those listed above; our maximum fitting-factors are above 0.93 for all waveforms \cite[cf.][]{Hotokezaka2013} and our minimum fitting-factors are above 0.90 across all waveforms~\cite[cf.][]{Tsang2019,Breschi2019}.  
    The fitting factor is more sensitive to deviations in the time-based phase or Fourier phase response, than it is to amplitude deviations. 
    The fits in \cite{Breschi2019} could possibly be improved by adding in more flexibility in the phase response.
    Our model bypasses the phase matching difficulty by directly fitting the phase with parameters, $(f_j, \alpha_j, \psi_j)$, from the injected signals from all three interferometers.
    Although \cite{Tsang2019} does directly fit the phase, the first-order model is too restricted to obtain higher fitting factors and better results may be obtained by increasing the order of the model.\par 
    
    Although numerical-relativity simulations currently provide the best estimate of the post-merger gravitational-wave strain, future post-merger signals may not be consistent with these state-of-the-art simulations.
    With this in mind, our model matches the numerical-relativity simulations well, but it is more flexible than these simulations.
    This is important because this method is a middle ground between  simulations of known waveforms, and more general (e.g. unmodelled excess power and \textsc{BayesWave}) methods. Nevertheless, numerical-relativity simulations are the primary method of investigating the dynamical physics of the post-merger region and  research into these simulations is vital.\par
    
\section{Acknowledgments}
        P.D.L. is supported through Australian Research Council (ARC) Future Fellowship FT160100112, ARC Discovery Project DP180103155, and ARC Centre of Excellence CE170100004. A.R.C. is supported by ARC grant DE190100656.
        We are grateful to Sukanta Bose for valuable comments on the manuscript.
\bibliography{FinalBib}

\pagebreak
\raggedbottom
\appendix
\renewcommand\thefigure{\thesection.\arabic{figure}} 
\renewcommand\thetable{\thesection.\arabic{table}} 
\setcounter{figure}{0} 
\setcounter{table}{0} 
\section{Numerical relativity simulations}
\label{appendix:a}
We use simulations from the \texttt{CoRe} gravitational wave database~\cite{Dietrich2018} for binary neutron star mergers. 
The simulations are listed by their equation of state, the progenitor mass, and the quadrupolar tidal deformability. 
We limited our simulations to those with equal-mass prognitors for compatibility with the heirarchical model in~\cite{Easter2019}. 
We chose simulations with the highest resolution such that the remnant was transmitting gravitational waves for $\sim 25$\,ms. 
In some cases increasing the resolution resulted in a reduced lifetime of the remnant. Table~\ref{tbl:NRwaveforms} shows the simulation designator for this paper, the name of the waveform in the \texttt{CoRe} database, and the citation for the associated simulation in the metadata (if available).
\begin{table}[H]
\centering
\caption{Numerical relativity simulations}\label{tbl:NRwaveforms}
 \begin{tabular}{|c|c|c|}
 \hline
 Designator & Simulation name & Citations \cite{Dietrich2018} \\  
 \hline
 \color{Waveform1}
 SLy-M1.350-$\Lambda$390 & \color{Waveform1} THC:0036:R03 & \color{Waveform1} \cite{Radice2016}  \\ 
 \hline
 \color{Waveform2}
 LS220-M1.350-$\Lambda$684 & \color{Waveform2} THC:0019:R05 & \color{Waveform2} \cite{radice2017} \\
 \hline
  \color{Waveform3}
 MS1b-M1.500-$\Lambda$864 & \color{Waveform3} BAM:0088:R01 & \color{Waveform3} - \\
 \hline
  \color{Waveform4}
  BHBlp-M1.300-$\Lambda$1046 & \color{Waveform4} THC:0002:R01 & \color{Waveform4} \cite{Radice2017a,Radice2018} \\
 \hline
  \color{Waveform5}
 DD2-M1.250-$\Lambda$1295 & \color{Waveform5} THC:0011:R01 &
   \color{Waveform5} \cite{Radice2017a,Radice2018} \\
   \hline
    \color{Waveform6}
  MS1b-M1.375-$\Lambda$1389 & \color{Waveform6} BAM:0070:R01 &
   \color{Waveform6} \cite{Dietrich2017} \\
   \hline
    \color{Waveform7}
   MS1b-M1.350-$\Lambda$1532 & \color{Waveform7} BAM:0065:R03 & \color{Waveform7}
  \cite{Bernuzzi2014} \\
  \hline
   \color{Waveform8}
  DD2-M1.200-$\Lambda$1612 & \color{Waveform8} THC:0010:R01 & 
  \color{Waveform8} \cite{Radice2017a,Radice2018} \\
  \hline
   \color{Waveform9}
  2H-M1.350-$\Lambda$2326 & \color{Waveform9} BAM:0002:R02 &
  \color{Waveform9} \cite{Bernuzzi2014}\\
 \hline
\end{tabular}
\end{table}
\section{Priors}
\label{appendix:b}
    The priors are listed in Eqs.~\ref{eq:firstprior}\Hyphdash*\ref{eq:w01} with $\mathcal{U}(a,b)$ representing a uniform prior distribution from $a$ to $b$. 
    The mode number $j$ is limited to $\lbrace 0,1,2 \rbrace$ and the mode number $i$ is restricted to $\lbrace 0,1 \rbrace$. 
    The priors in Eqs.~\ref{eq:modesorting}\Hyphdash*\ref{eq:w01} are constrained priors. 
    These restrictions are enforced in addition to the standard priors. The prior in Eq.~\ref{eq:modesorting} ensures that the maximum spectral amplitude of each mode is decreasing. 
    This results in $f_0$ converging to the loudest peak. 
    \begin{eqnarray}
        \log_{10}{H} & \sim & \mathcal{U}(-24, -19)\label{eq:firstprior}  \\
        f_j & \sim & \mathcal{U}(1000,5000) \\
        \log_{10}T_j & \sim & \mathcal{U}(-4.0,0.3)\\
        \psi_j & \sim & \mathcal{U}(-\pi,\pi) \\
        \alpha_j & \sim & \mathcal{U}(-6.4,6.4) \\
        w_i & \sim & \mathcal{U}(0.0,1.0)\label{eq:lastprior}\\
        \log_{10}\left(\frac{\max|\tilde{h}_{j}(f)|_f}
        {\max|\tilde{h}_{j+1}(f)|_f}\right) & \sim & \mathcal{U}(0.0,10.0)\label{eq:modesorting} \\
        w_0+w_1 & \sim & \mathcal{U}(0.0,1.0) \label{eq:w01}
    \end{eqnarray}
    $w_2$ is calculated as:
    \begin{equation}
        w_2 = 1 - w_0 - w_1,
    \end{equation}
    ensuring that $\sum_j w_j = 1$ and $w_2\in[0,1]$ as required. 
% \pagebreak
\section{Posteriors for all numerical-relativity injections}
\label{appendix:c}  
\begin{figure}[H]
        \centering
        \includegraphics[scale=0.37]{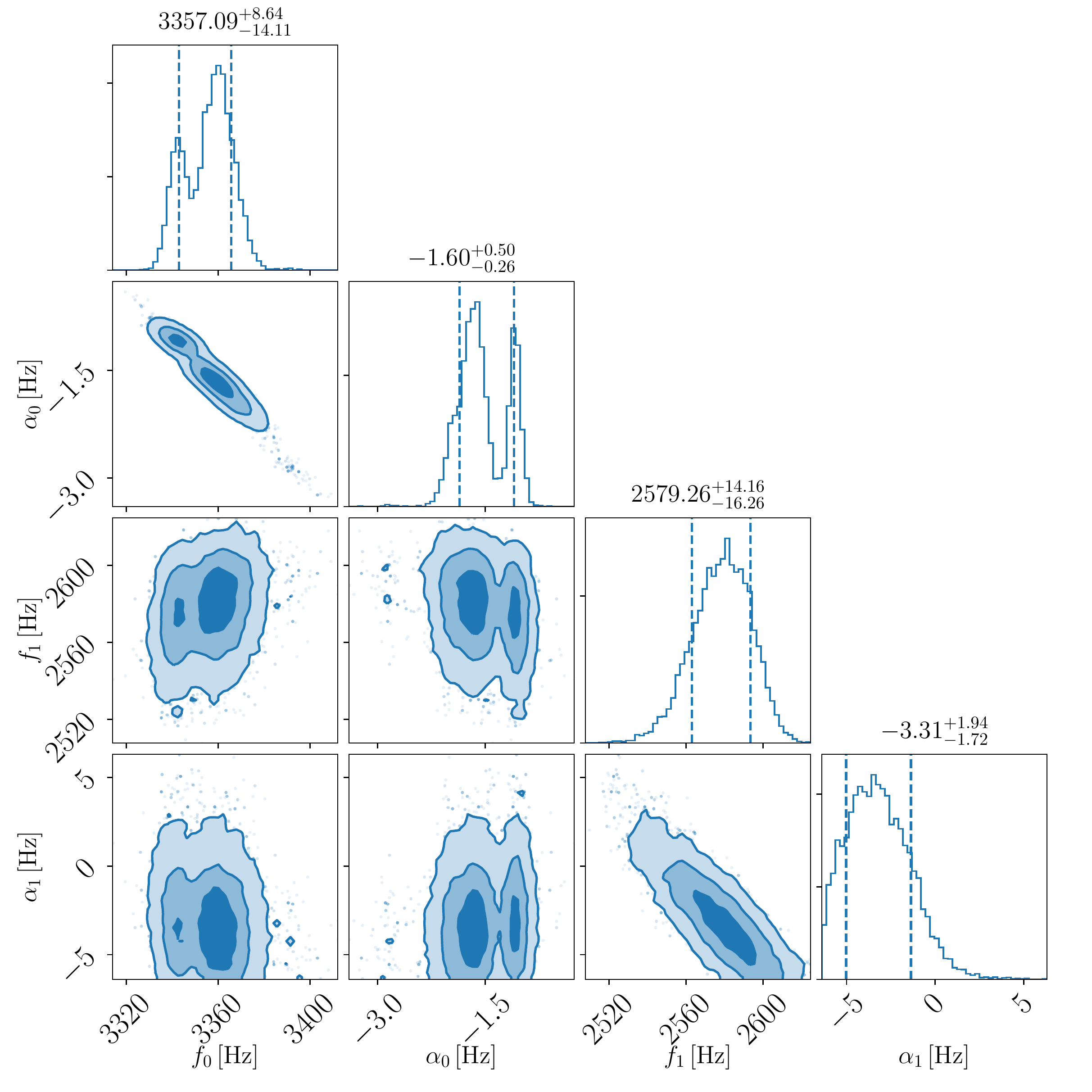}
        \caption
        {Selected posteriors for numerical-relativity post-merger injection using the equation of state SLy with equal mass, $1.35\,\mathrm{M}_\odot$, neutron stars (waveform SLy-M1.350-$\Lambda$390). The numerical-relativity simulation was injected at a post-merger signal-to-noise ratio of 50.} 
        \label{fig:Corner1}
\end{figure}

\begin{figure}[H]
        \centering
        \includegraphics[scale=0.37]{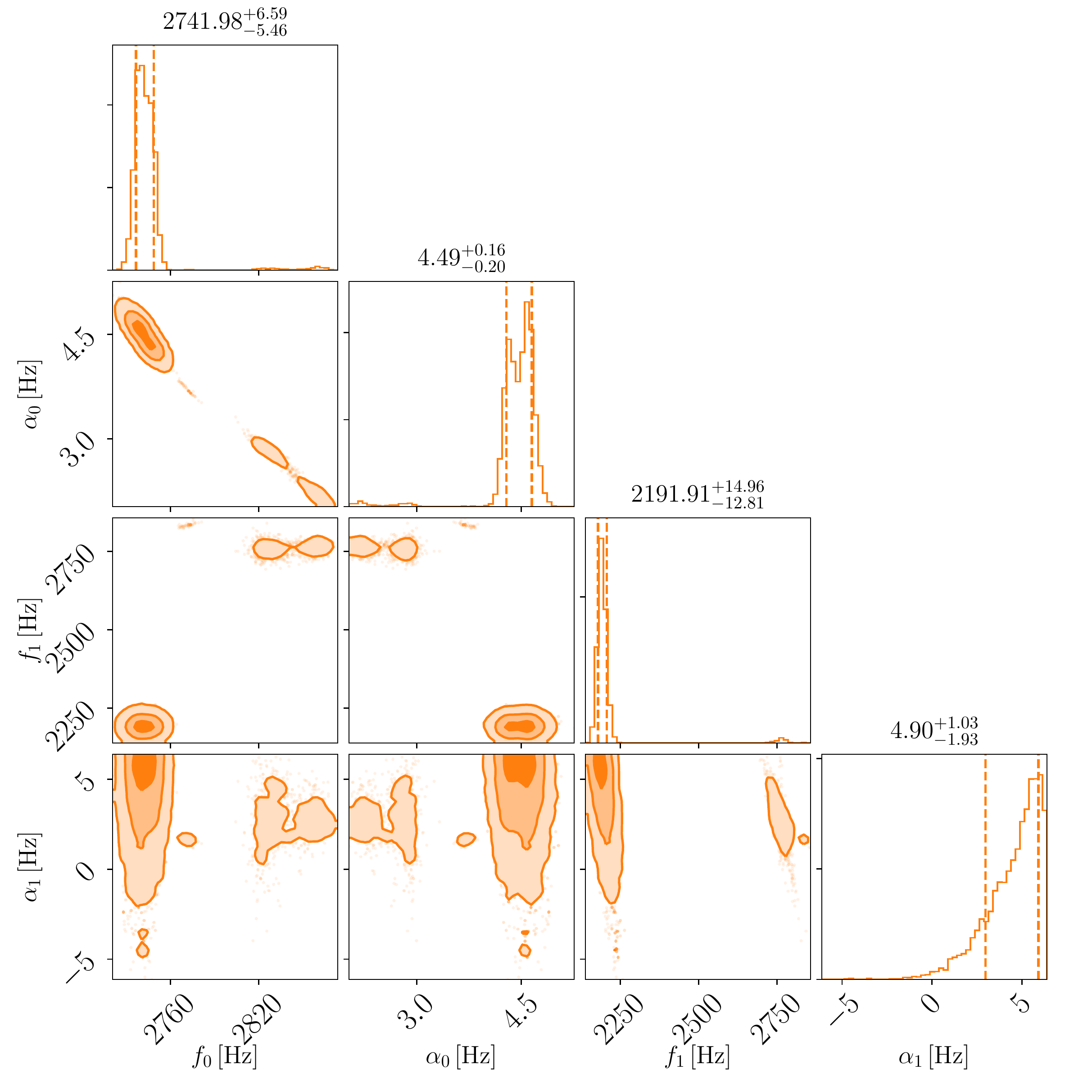}
        \caption
        {As per Fig.~\ref{fig:Corner1} using the equation of state LS220 with equal mass, $1.35\,\mathrm{M}_\odot$, neutron stars (waveform LS220-M1.350-$\Lambda$684).} 
        \label{fig:Corner2}
\end{figure}

\begin{figure}[H]
        \centering
        \includegraphics[scale=0.37]{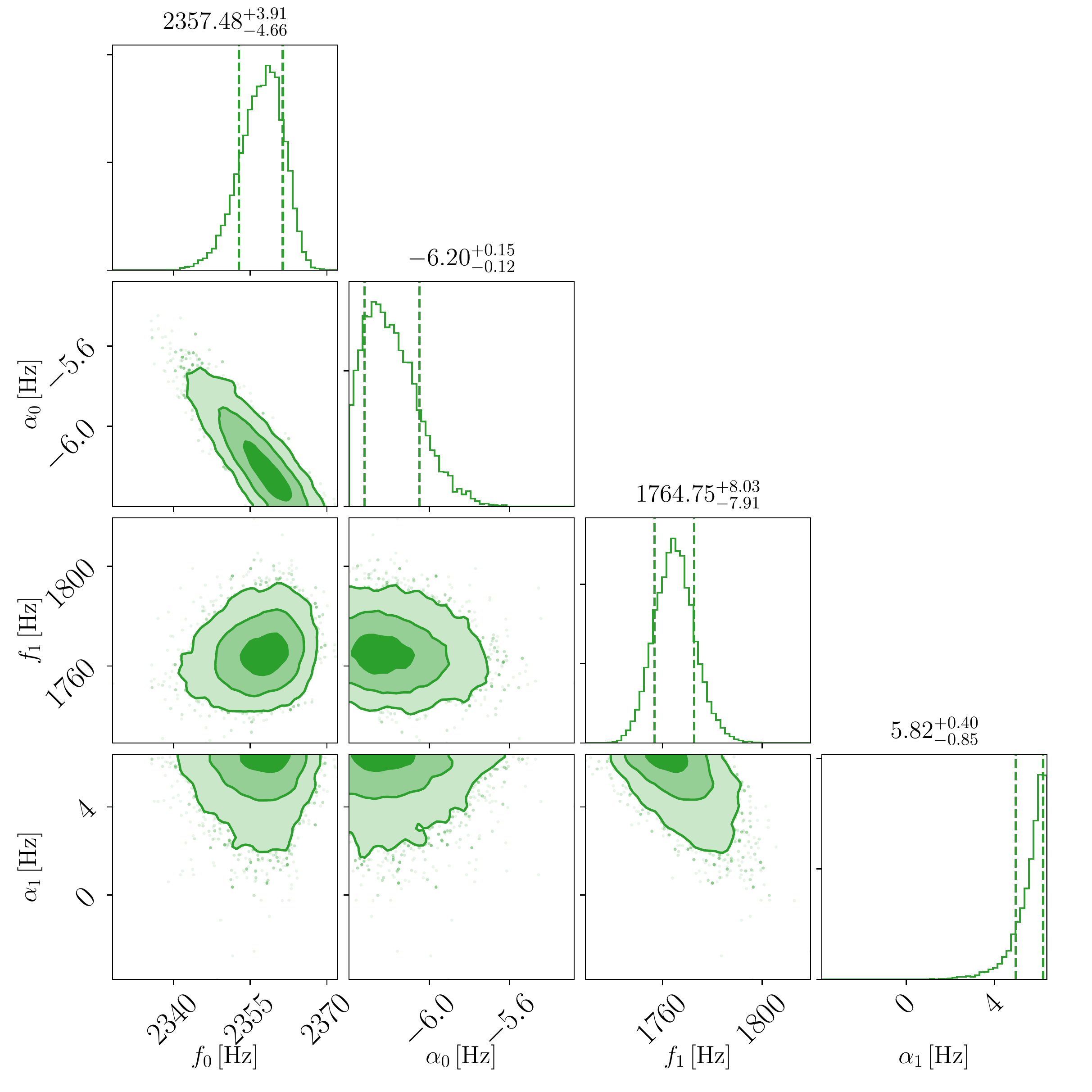}
        \caption
        {As per Fig.~\ref{fig:Corner1} using the equation of state MS1b with equal mass, $1.50\,\mathrm{M}_\odot$, neutron stars (waveform MS1b-M1.500-$\Lambda$864).} 
        \label{fig:Corner3}
\end{figure}

\begin{figure}[H]
        \centering
        \includegraphics[scale=0.37]{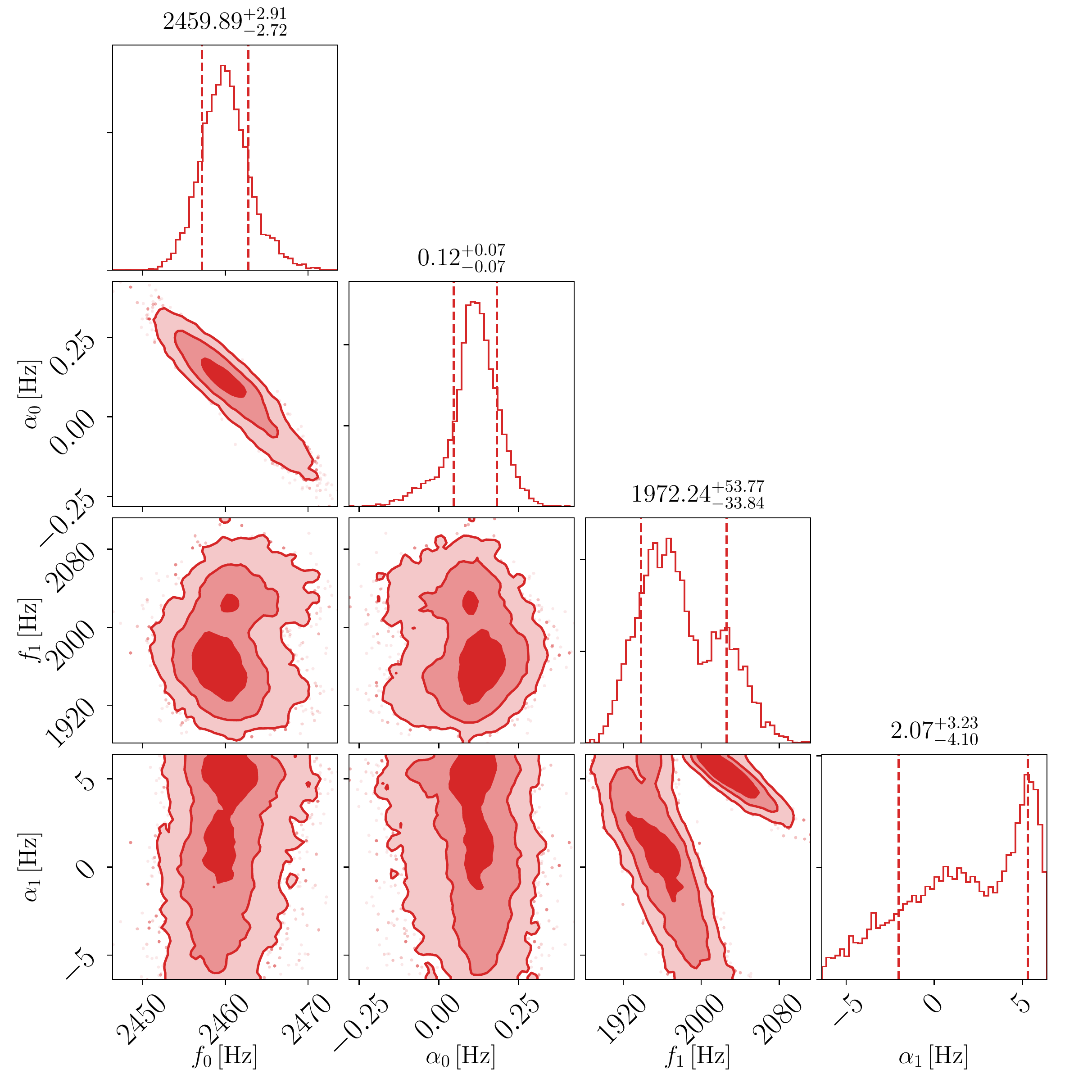}
        \caption
        {As per Fig.~\ref{fig:Corner1} using the equation of state BHBlp with equal mass, $1.30\,\mathrm{M}_\odot$, neutron stars (waveform BHBlp-M1.300-$\Lambda$1046).} 
        \label{fig:Corner4}
\end{figure}

\begin{figure}[H]
        \centering
        \includegraphics[scale=0.37]{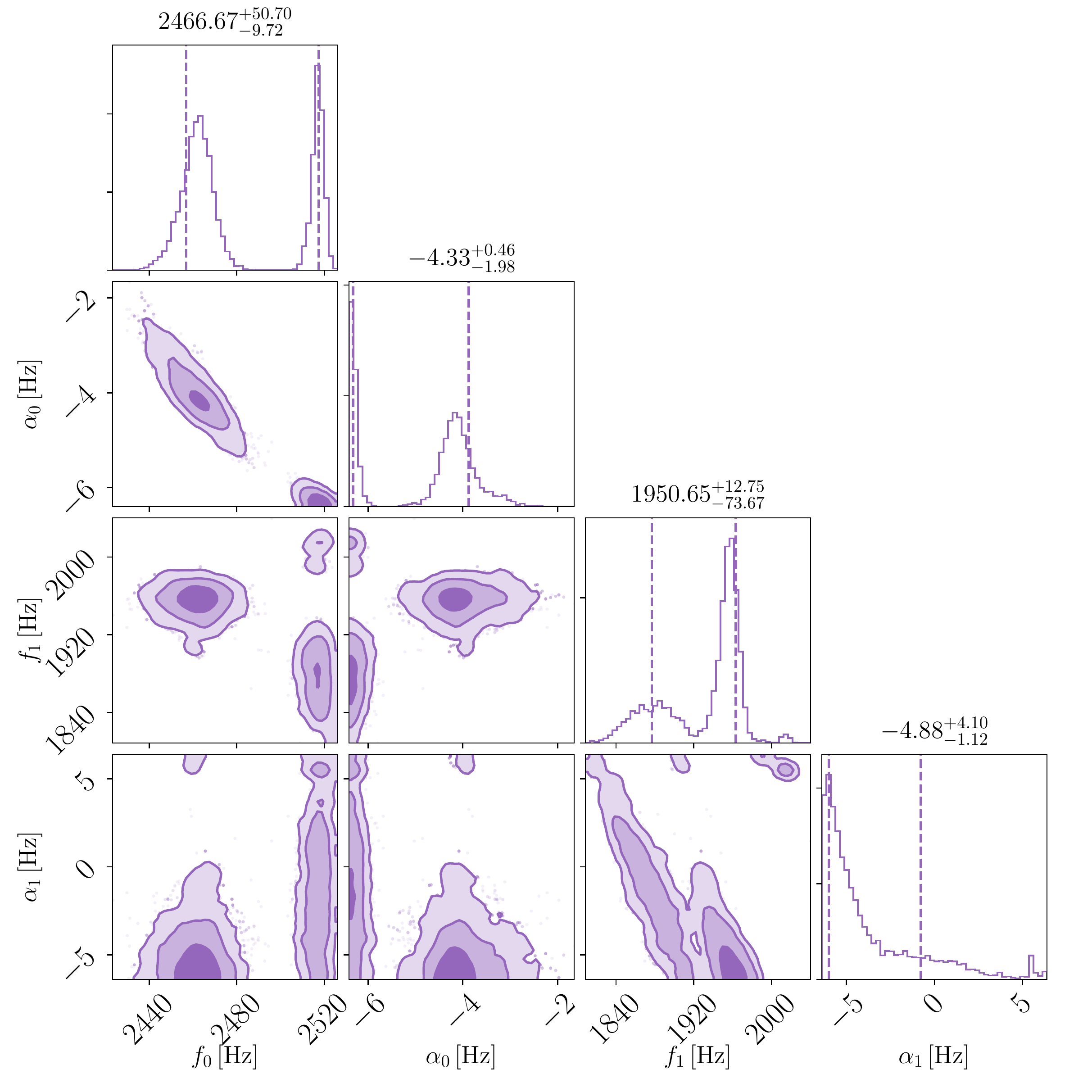}
        \caption
        {As per Fig.~\ref{fig:Corner1} using the equation of state DD2 with equal mass, $1.25\,\mathrm{M}_\odot$, neutron stars (waveform DD2-M1.250-$\Lambda$1295).} 
        \label{fig:Corner5}
\end{figure}

\begin{figure}[H]
        \centering
        \includegraphics[scale=0.37]{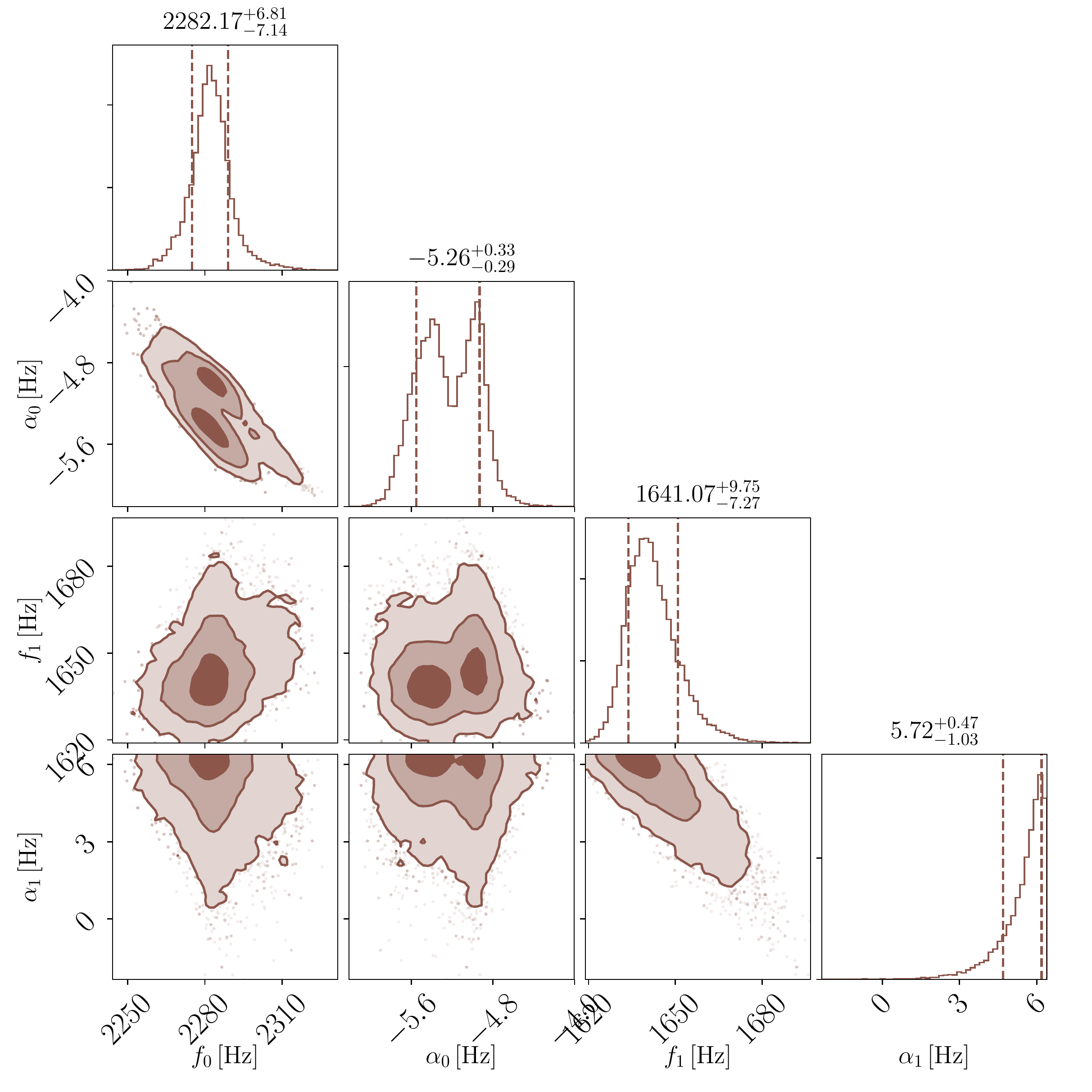}
        \caption
        {As per Fig.~\ref{fig:Corner1} using the equation of state MS1b with equal mass, $1.375\,\mathrm{M}_\odot$, neutron stars (waveform MS1b-M1.375-$\Lambda$1389).}
        \label{fig:Corner6}
\end{figure}

\begin{figure}[H]
        \centering
        \includegraphics[scale=0.37]{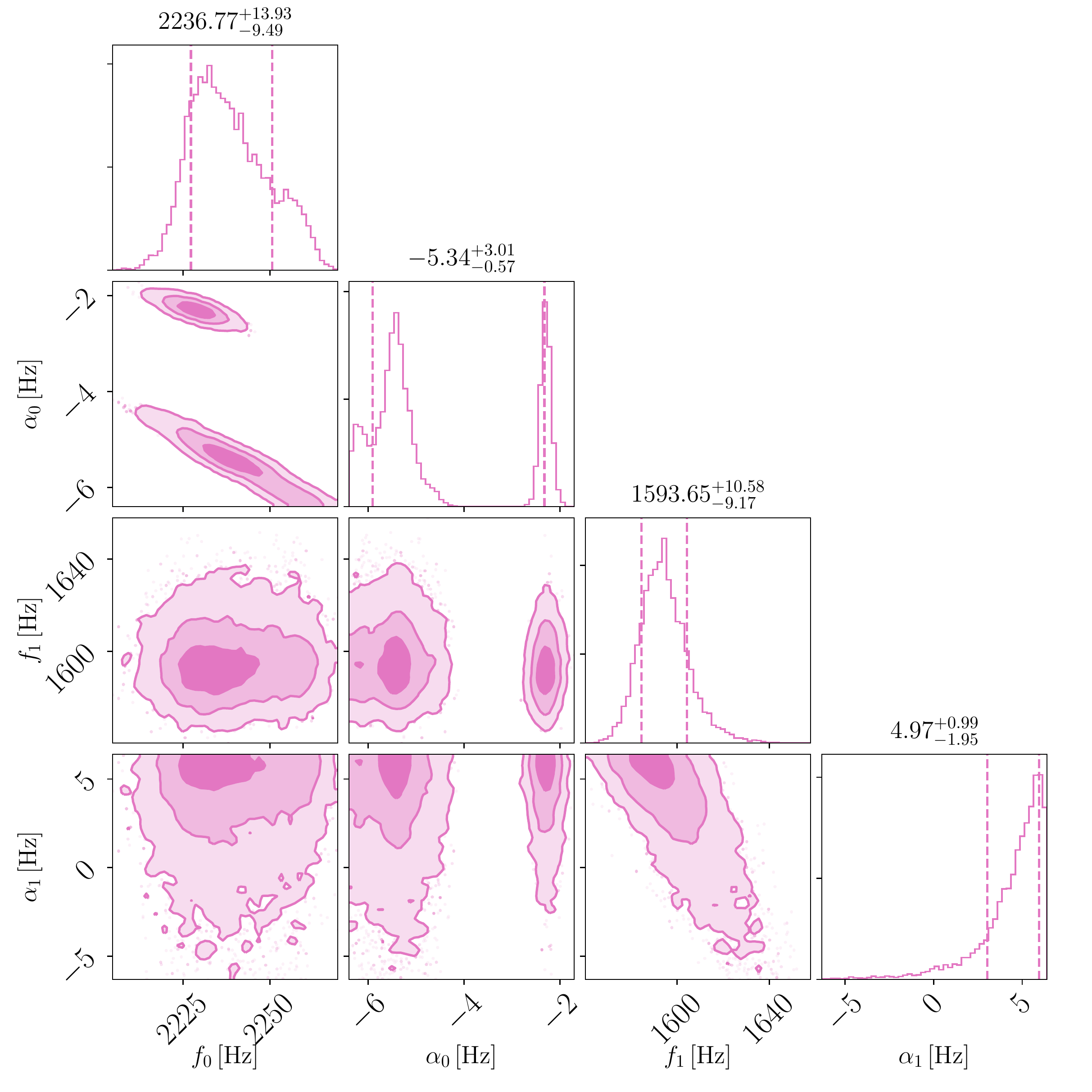}
        \caption
        {As per Fig.~\ref{fig:Corner1} using the equation of state MS1b with equal mass, $1.35\,\mathrm{M}_\odot$, neutron stars (waveform MS1b-M1.350-$\Lambda$1532).}
        \label{fig:Corner7}
\end{figure}

\begin{figure}[H]
        \centering
        \includegraphics[scale=0.37]{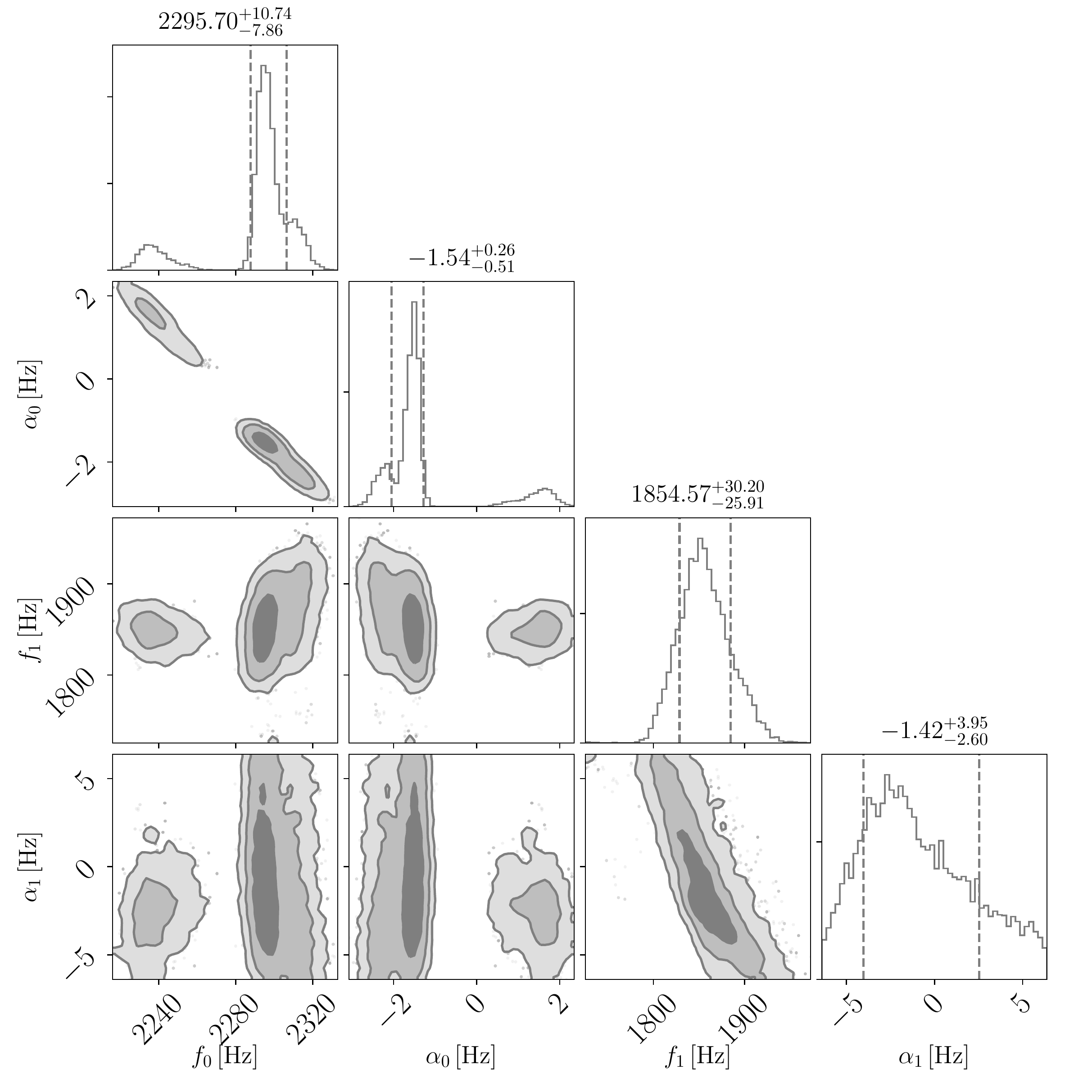}
        \caption
        {As per Fig.~\ref{fig:Corner1} using the equation of state DD2 with equal mass, $1.20\,\mathrm{M}_\odot$, neutron stars (waveform DD2-M1.200-$\Lambda$1612).}
        \label{fig:Corner8}
\end{figure}

\begin{figure}[H]
        \centering
        \includegraphics[scale=0.37]{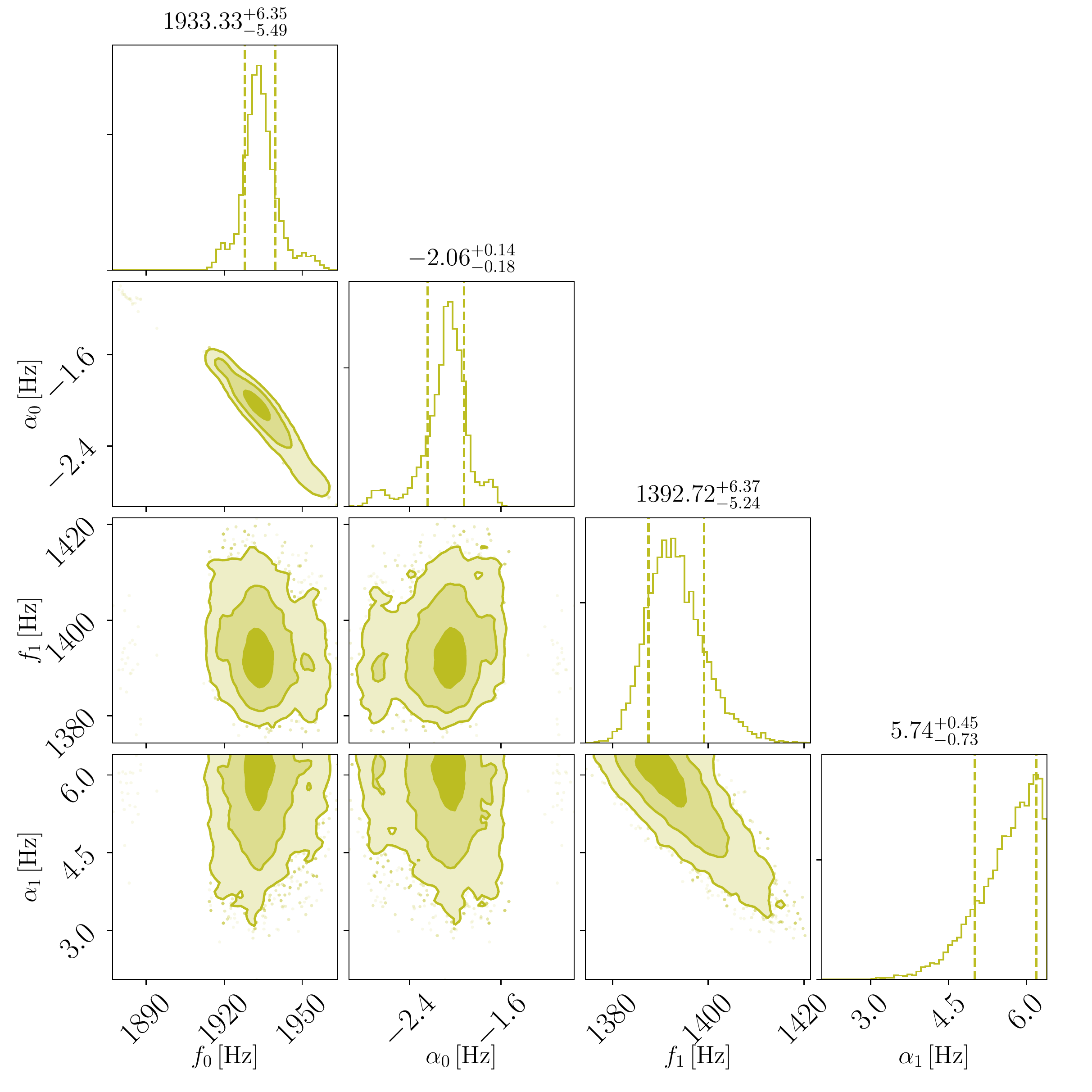}
        \caption
        {As per Fig.~\ref{fig:Corner1} using the equation of state 2H with equal mass, $1.35\,\mathrm{M}_\odot$, neutron stars (waveform 2H-M1.350-$\Lambda$2326).}
        \label{fig:Corner9}
\end{figure}

Selected posteriors for all numerical-relativity simulations are shown in Fig.~\ref{fig:Corner1}-\ref{fig:Corner9}. 
The waveforms are injected at a post-merger signal-to-noise ratio of 50.
The posteriors shown are: $f_0, \alpha_0, f_1$ and $\alpha_1$. 
The posteriors are coloured as per Fig.~\ref{fig:FittingFactors} and Table~\ref{tbl:NRwaveforms}.
\newpage


\end{document}